\documentclass{aa}
\usepackage{longtable}
\usepackage{supertabular}
\usepackage{graphics}
\usepackage{rotating}
\usepackage{txfonts}
\usepackage{natbib}
\begin{document}
\title{RASS-SDSS Galaxy Cluster Survey. V.} 
\subtitle{The X-ray-Underluminous Abell Clusters.}
\author{P. Popesso\inst{1}, A. Biviano\inst{2}, H. B\"ohringer\inst{3}, M. Romaniello\inst{1}}
\institute{ European Southern Observatory, Karl Scharzschild Strasse 2, D-85748
\and INAF - Osservatorio Astronomico di Trieste, via G. B. Tiepolo 11, I-34131, Trieste, Italy
\and Max-Planck-Institut fur extraterrestrische Physik, 85748 Garching, Germany}

\abstract
{In this paper we consider a large sample of optically selected
clusters, in order to elucidate the physical reasons for the existence
of X-ray underluminous clusters. For this purpose we analyze the
correlations of the X-ray and optical properties of a sample of 137
spectroscopically confirmed Abell clusters in the SDSS database.  We
search for the X-ray counterpart of each cluster in the ROSAT All Sky
Survey. We find that 40\% of our clusters have a marginal X-ray
detection or remain undetected in X-rays. These clusters appear too
X-ray faint on average for their velocity disperiosn determined mass,
i.e. they do not follow the scaling relation between X-ray luminosity
and virial mass traced by the other clusters. On the other hand, they
do follow the general scaling relation between optical luminosity and
virial mass. We refer to these clusters as the X-ray-Underluminous
Abell clusters (AXU clusters, for short) and designate as 'normal' the
X-ray detected Abell systems. We examine the distributions and
properties of the galaxy populations of the normal and the AXU
clusters, separately. The AXU clusters are characterized by
leptokurtic (more centrally concentrated than a Gaussian) velocity
distribution of their member galaxies in the outskirts ($1.5 <
r/r_{200} \leq 3.5$), as expected for the systems in accretion. In
addition, the AXU clusters have a higher fraction of blue galaxies in
the external region and show a marginally significant paucity of
galaxies at the center. Our results seem to support the interpretation
that the AXU clusters are systems in formation undergoing a phase of
mass accretion. Their low X-ray luminosity should be due to the still
accreting Intracluster gas or to an ongoing merging process.}
\authorrunning{P. Popesso et al.}
\maketitle

\section{Introduction}
Clusters of galaxies are extremely important astrophysical tools. They
are the most massive gravitationally bound systems in the
universe. Since they sample the high mass end of the mass function of
collapsed systems, they can be used to provide tight constraints on
cosmological parameters such as $\Omega_m$, $\sigma_8$ and
$\Lambda$ (Eke at al 1996, Donahue \& Voit 1999). Moreover they are
extremely powerful laboratories to study galaxy formation and
evolution. To investigate the global properties of the cosmological
background it is necessary to construct and study a large sample of
clusters (Borgani \& Guzzo 2001). 

Several techniques exist to build cluster samples, each based on
different clusters properties. The first attempts at a large,
homogeneous survey for galaxy clusters was conducted by Abell (1958)
with the visual identification of clusters on the Palomar Observatory
Sky Survey (POSS) photographic plates. Similar cataloges were
constructed by Zwicky and collaborators (Zwicky et al. 1968). Since
then, a large number of optically selected samples have been
constructed with automated methods: EDCC (Edimburgh Durham Cluster
Catalogue: Lumdsen et al. 1992), APM (Automatic Plate measuring;
Dalton et al. 1994), PSCS (Palomar Distant Cluster Survey; Postman et
al. 1996), EIS (ESO Imaging Cluster Survey; Olsen et al. 1999),
ENACS (ESO Nearby Abell Cluster Survey, Katgert et al. 1996,
Mazure et al. 1996), RCS (Red sequence Cluster Survey; Gladders \&
Yee 2000) and the samples derived from the Sloan Digital Sky Survey
(Goto at al. 2002; Bahcall et al. 2003). The advantage of using
optical data is that in general it is relatively easy to build large
optically selected cluster catalogs, which allow one to investigate
cluster properties with a statistically solid data-base. On the
other hand, the main disadvantage of the optical selection is that the
selection procedure can be seriously affected by projection
effects. Only a very observationally expensive spectroscopic campaign
can confirm the overdensities in 3 dimensions.

In 1978, the launch of the first X-ray imaging telescope, the
\emph{Einstein} observatory, began a new era of cluster discovery, as
clusters proved to be luminous ($\ge 10^{42-45}$ ergs $\rm{s}^{-1}$),
extended ($\rm{r} \ga 1$ Mpc) X-ray sources, readily identified in the
X-ray sky.  Therefore, X-ray observations of galaxy clusters provided
an efficient and physically motivated method of identification of
these structures.  The X-ray selection is more robust against
contamination along the line-of-sight than traditional optical
methods, because the X-ray emission, unlike galaxy overdensities, is
proportional to the square of the (gas) density. The ROSAT satellite
with its large field of view and better sensitivity, allowed to a leap
forward in the X-ray cluster astronomy, producing large samples of
both nearby and distant clusters (Castander et al. 1995; Ebeling et
al. 1996a, 1996b; Scharf et al. 1997; Ebeling et al. 2000; B\"ohringer
et al. 2001; Gioia et al. 2001; B\"ohringer et al. 2002; Rosati et
al. 2002 and references therein). The disadvantage of X-ray cluster
surveys is their lower efficiency and higher observational cost as
compared to optical surveys.

It is clear that understanding the selection effects and the biases
due to the different cluster selection techniques is crucial for
interpreting the scientific results obtained from such different
cluster samples. Castander et al. (1994) used ROSAT to observe cluster
candidates in the redshift range 0.7-0.9 from the 3.5 square degree
subsample of Gunn et al.'s (1986) optical cluster catalog and found
surprisingly weak X-ray emission. Bower et al. (1994) undertook ROSAT
X-ray observations of optically selected clusters from Couch et al.'s
(1991) 46 $\rm{deg}^2$ catalog. Bower et al. (1994) selected a random
subset of the full catalogue, in the redshift range 0.15--0.66.  The
X-ray luminosity of almost all the selected clusters was found to be
surprisingly low, suggesting, on the one hand, substantial evolution
of the X-ray luminosity function between redshift $z=0$ and $z \sim
0.4$, and, on the other hand, overestimated velocity dispersions for
the nearby X-ray underluminous clusters, perhaps as a consequence of
the contamination by galaxy filaments and of radial infall of field
galaxies into the clusters.  Similar results were obtained by Holden
et al. (1997).

With the ROSAT Optical X-ray Survey (ROXS), Donahue et al. (2002)
concluded that there is little overlap between the samples of
X-ray-selected and optically-selected galaxy clusters.  Only $\sim
20$\% of the optically selected clusters were found in X-rays, while
$\sim 60$\% of the X-ray clusters were also identified in the optical
sample. Furthermore, not all of the X-ray detected clusters had a
prominent red-sequence, something that could introduce a selection
bias in those cluster surveys based on colour information (Goto et
al. 2002, Gladders \& Yee 2000). Ledlow at al. (2003) analyzed the
X-ray properties of a sample of nearby bright Abell clusters, using
the ROSAT All-Sky Survey (RASS). They found an X-ray detection rate of
83\%. Gilbank at al. (2004) explored the biases due to optical and
X-ray cluster selection techniques in the X-ray Dark Cluster Survey
(XDCS). They found that a considerable fraction of the optically
selected clusters do not have a clear X-ray counterpart, yet
spectroscopic follow-up of a subsample of X-ray underluminous systems
confirmed their physical reality. Lubin et al. (2004) analyzed the
X-ray properties of two optically selected clusters at $z \ge 0.7$,
with XMM-Newton. They found the two clusters are characterized by
X-ray luminosities and temperatures that are too small for their
measured velocity dispersion. Similar results were obtained in the
XMM-2dF Survey of Basilakos et al. (2004). They found many more
optical cluster candidates than X-ray ones. Deeper XMM data confirmed
that their X-ray undetected cluster candidates have intrinsically
very low X-ray luminosities.

In this paper we consider a large sample of optically- and
X-ray-selected clusters, in order to elucidate the physical reasons
for the existence of underluminous optical/X-ray clusters.  The
starting point of this work is the analysis we conducted on a sample
of X-ray selected clusters sample (Popesso et 2005a, Paper III of this
series).  90\% of those systems are taken from the REFLEX and NORAS
catalogs, which are X-ray flux-limited cluster catalogs entirely built
upon the ROSAT-All-Sky Survey (RASS). The remaining 10\% of that
sample are groups or faint clusters with X-ray fluxes below the flux
limits of REFLEX and NORAS. In Paper III  we found an
optical counterpart for each of the X-ray selected clusters of the
RASS. Using Sloan Digital Sky Survey (SDSS, see, e.g., Abazajian et
al.  2003) optical data for these clusters, we then studied the
scatter of the correlations between several optical and X-ray cluster
properties (X-ray and optical luminosities, mass, velocity dispersion
and temperature). In this paper we extend our analysis to a sample of
{\em optically} selected clusters.

The paper is organized as follows. In section 2 we describe the data
and the sample of optically selected clusters used for the
analysis. We also describe how we measure the optical
luminosity, the velocity dispersion, the mass and the X-ray luminosity
of the clusters. In section 3 we analyze the correlation of both the
X-ray and the optical cluster luminosities with their masses. In
section 4 we describe the optical properties of the Abell clusters
without clear X-ray detection and compare them with those of normal
X-ray emitting Abell systems. In section 6 we discuss our results
and give our conclusions.

We adopt a Hubble constant $\rm{H}_0=70 \; \rm{h} \; \rm{km} \;
\rm{s}^{-1} \; \rm{Mpc}^{-1}$, and a flat geometry of the Universe,
with $\Omega_{m}=0.3$ and $\Omega_{\Lambda}=0.7$ throughout this paper.

\section{The data}\label{s-data}
The optical data used in this paper are taken from the SDSS (Fukugita
et al. 1996, Gunn et al. 1998, Lupton et al. 1999, York et al. 2000,
Hogg et al. 2001, Eisenstein et al. 2001, Smith et al. 2002, Strauss
et al. 2002, Stoughton et al.  2002, Blanton et al. 2003 and Abazajian
et al.  2003).  The SDSS consists of an imaging survey of $\pi$
steradians of the northern sky in the five passbands $u, g, r ,i, z,$
covering the entire optical range.  The imaging survey is taken in
drift-scan mode.  The imaging data are processed with a photometric
pipeline specially written for the SDSS data (PHOTO, Lupton et
al. 2001).  For each cluster we defined a photometric galaxy catalog
as described in Section 3 of Popesso et al. (2004, Paper I of this
series, see also Yasuda et al. 2001).  For the analysis in this paper
we use only SDSS Model magnitudes.

The spectroscopic component of the survey is carried out using two
fiber-fed double spectrographs, covering the wavelength range
3800--9200 \AA, over 4098 pixels. They have a resolution
$\Delta\lambda/\lambda$ varying between 1850 and 2200, and together
they are fed by 640 fibers, each with an entrance diameter of 3
arcsec.  

The X-ray data are taken from the RASS. The RASS was conducted mainly
during the first half year of the ROSAT mission in 1990 and 1991
(Tr\"umper 1988). The ROSAT mirror system and the Position Sensitive
Proportional counter (PSPC) operating in the soft X-ray regime
(0.1-2.4 keV) provided optimal conditions for the studies of celestial
objects with low surface brightness. In particular, due to the
unlimited field of view of the RASS and the low background of the
PSPC. This dataset is ideal to investigate the properties of nearby
clusters of galaxies.

\subsection{The cluster samples}\label{s-sample}

\subsubsection{The X-ray selected cluster sample}
\begin{figure}
\begin{center}
\begin{minipage}{0.5\textwidth}
\resizebox{\hsize}{!}{\includegraphics{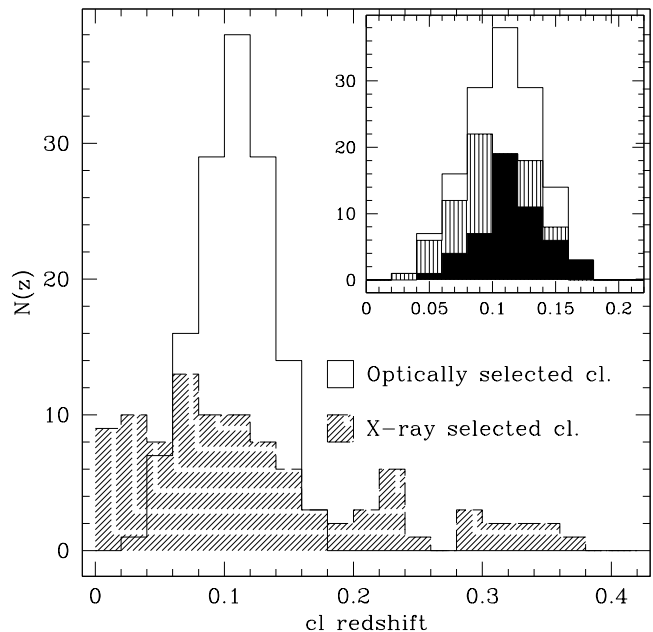}}
\end{minipage}
\end{center}
\caption{
Redshift distribution of the X-ary and optically selected,
rispectively, cluster samples used in this paper. The solid line in
the main panel shows the redshift distribution of the optically
selected cluster sample and the shaded histogram is the redshift
distribution of the X-ray clusters. The small panel in the figure
shows the redshift distribution of the X-ray detected (grey histogram)
and the X-ray undetected (black histogram) optically selected
clusters. The solid line in the small panel shows the redshift
distribution of the whole optically selected cluster sample for
comparison. }
\label{z_dist}
\end{figure}

As reference X-ray cluster sample for the comparison between
X-ray and optically selected clusters, we consider a subsample of the
X-ray selected RASS-SDSS Galaxy Cluster Sample of Popesso et
al. (2005b). The RASS-SDSS galaxy cluster catalog comprises 130
systems detected in the ROSAT All Sky Survey (RASS).  The X-ray
cluster properties and the redshifts have been taken from different
catalogs of X-ray selected clusters: the ROSAT-ESO flux limited X-ray
cluster sample (REFLEX, B\"ohringer et al.  2001, 2002), the Northern
ROSAT All-sky cluster sample (NORAS, B\"ohringer et al.  2000), the
NORAS 2 cluster sample (Retzlaff 2001), the ASCA Cluster Catalog (ACC)
from Horner et al. (2001) and the Group Sample (GS) of Mulchaey et
al. (2003). The subsample considered in this paper comprises the
RASS-SDSS galaxy clusters with known mass (either the virial estimate
from optical data, or, when this is not available, the mass derived
from the X-ray temperature) for a total number of 102 systems (69
cluster with known optical mass $+$ 33 clusters with mass derived from
the mass-temperature realation). The sample is drwan from the SDSS DR2
imaging data which cover 3324 square degrees. The considered cluster
sample covers the entire range of masses and X-ray/optical
luminosities, from very low-mass and X-ray/optical faint groups
($10^{13} M{\odot}$) to very massive and X-ray/optical bright clusters
($5\times10^{15} M{\odot}$). The cluster sample comprises mainly
nearby systems at the mean redshift of 0.1 and few objects (10) in the
range $0.25 \le z \le 0.37$. The redshift distribution of the cluster
sample is shown in Fig. \ref{z_dist}.

\subsubsection{The optically selected cluster sample}
The optically selected cluster sample considered in this paper is a
subsample of the Abell cluster catalog (Abell 1958). We have selected
all the Abell clusters in the region covered by the $3^{rd}$ data
release (DR3) of the SDSS (5282 $\rm{deg}^{-2}$). The Abell catalog is
based on a visual inspection of galaxy overdensities. Therefore, it is
affected by the presence of spurious detections due to projection
effects. To exclude the spurious clusters from the catalog, we
consider only the clusters with a spectroscopic confirmation of the
galaxy overdensity. For this, we use the SDSS spectroscopic catalog,
which provides spectra and redshifts for more than 250,000 galaxies
with Petrosian magnitude $r_{Petro} \le 17.77$.

We estimate the mean cluster spectroscopic redshift $z_c$ as the peak
of the overdensity in the redshift distribution of the galaxies around
the cluster coordinates. Since the purpose of this paper is to compare
optical and X-ray properties of galaxy clusters, it is extremely
important to avoid misclassification between the optical and the X-ray
sources. Therefore, we have checked our estimations of the mean
cluster redshift with those available in the literature, as well
as with the photometric $z_c$ estimate obtained from the relation that
links the mean redshift of a cluster with the apparent magnitude of
its tenth brightest galaxy (Abell et al. 1989).  Clusters for which
discrepancies were found among the different $z_c$ estimates are
excluded from the final sample used in this paper.

\begin{figure*}
\begin{center}
\begin{minipage}{0.36\textwidth}
\resizebox{\hsize}{!}{\includegraphics{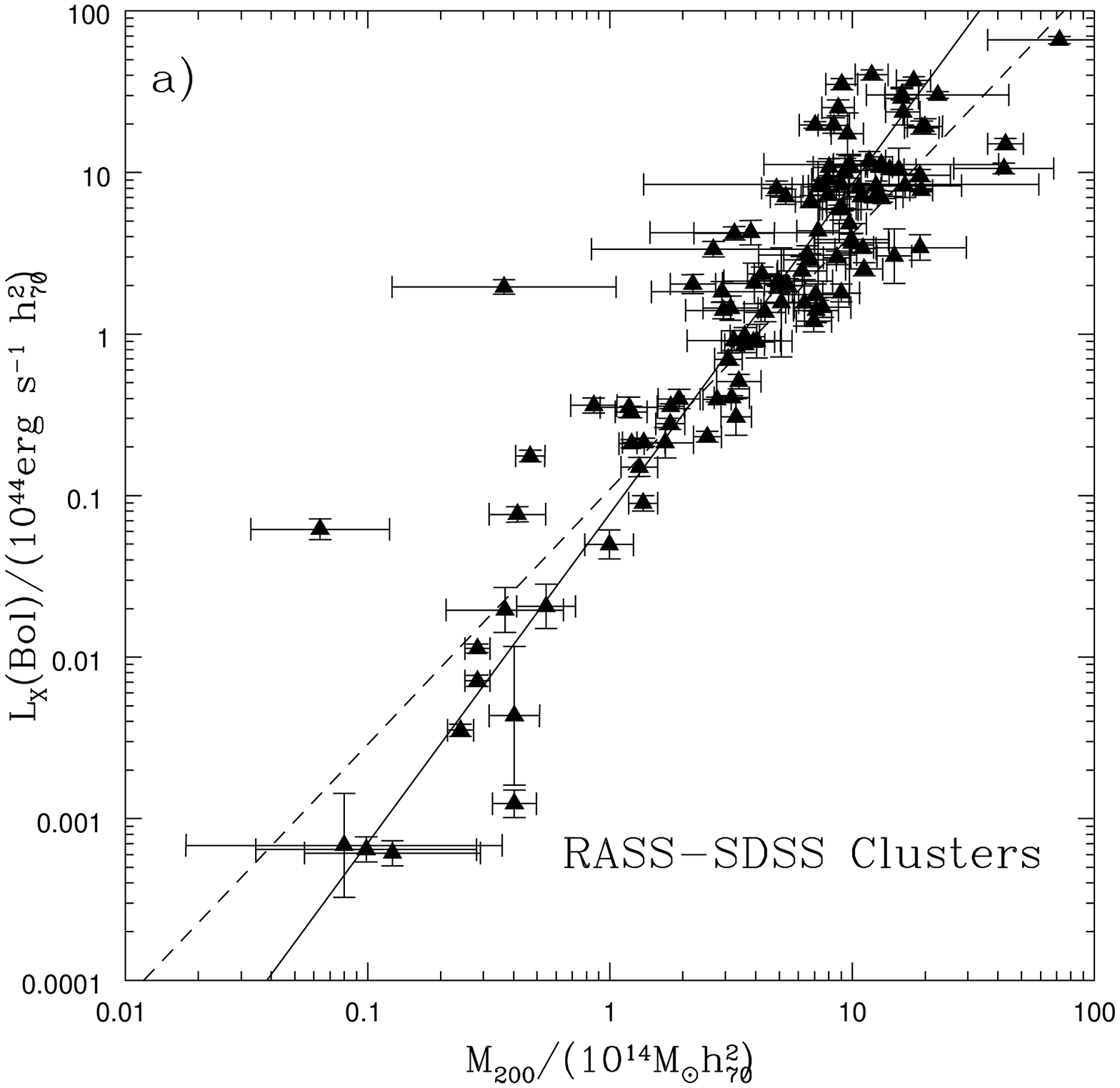}}
\end{minipage}
\begin{minipage}{0.36\textwidth}
\resizebox{\hsize}{!}{\includegraphics{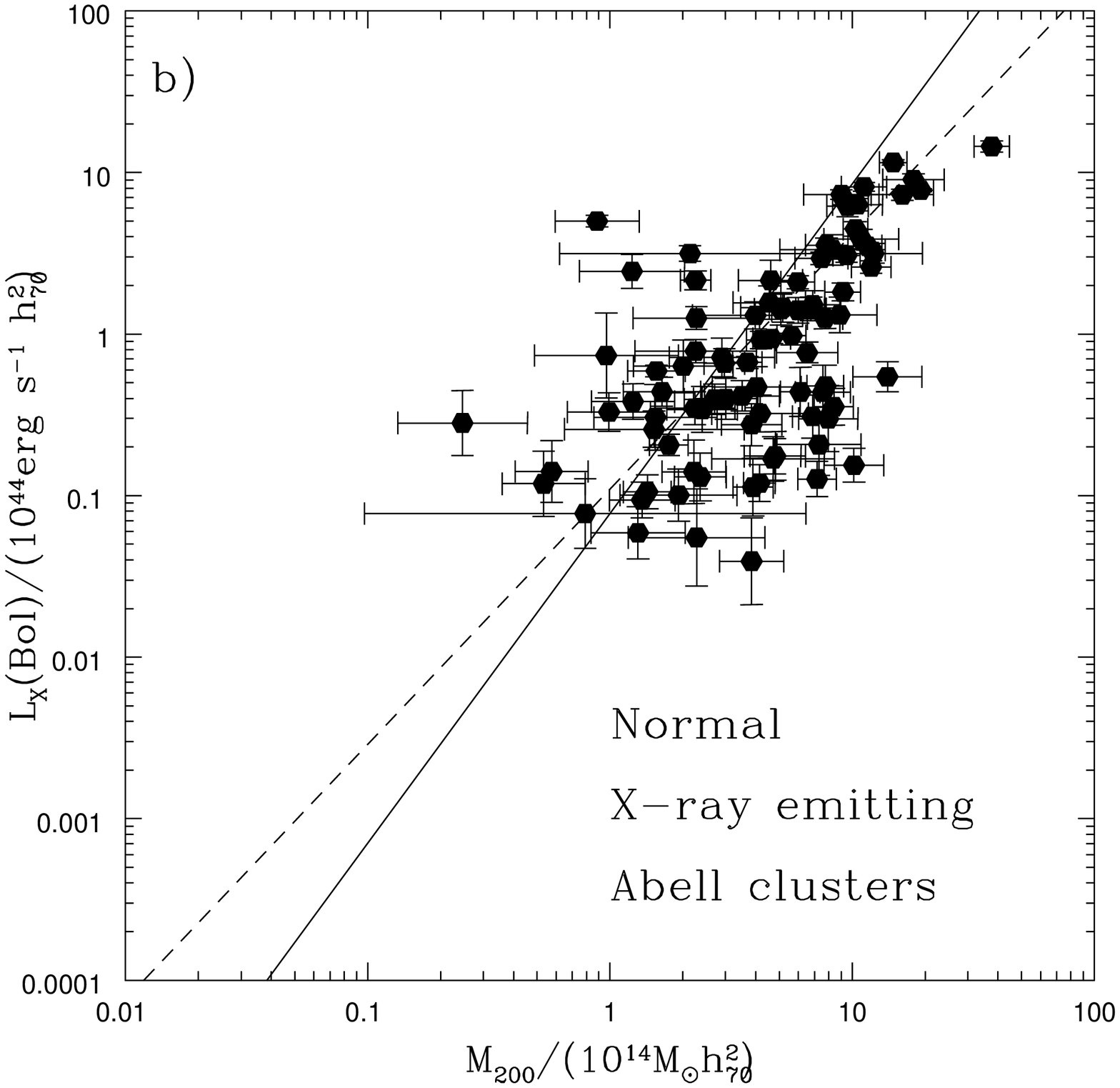}}
\end{minipage}
\begin{minipage}{0.36\textwidth}
\resizebox{\hsize}{!}{\includegraphics{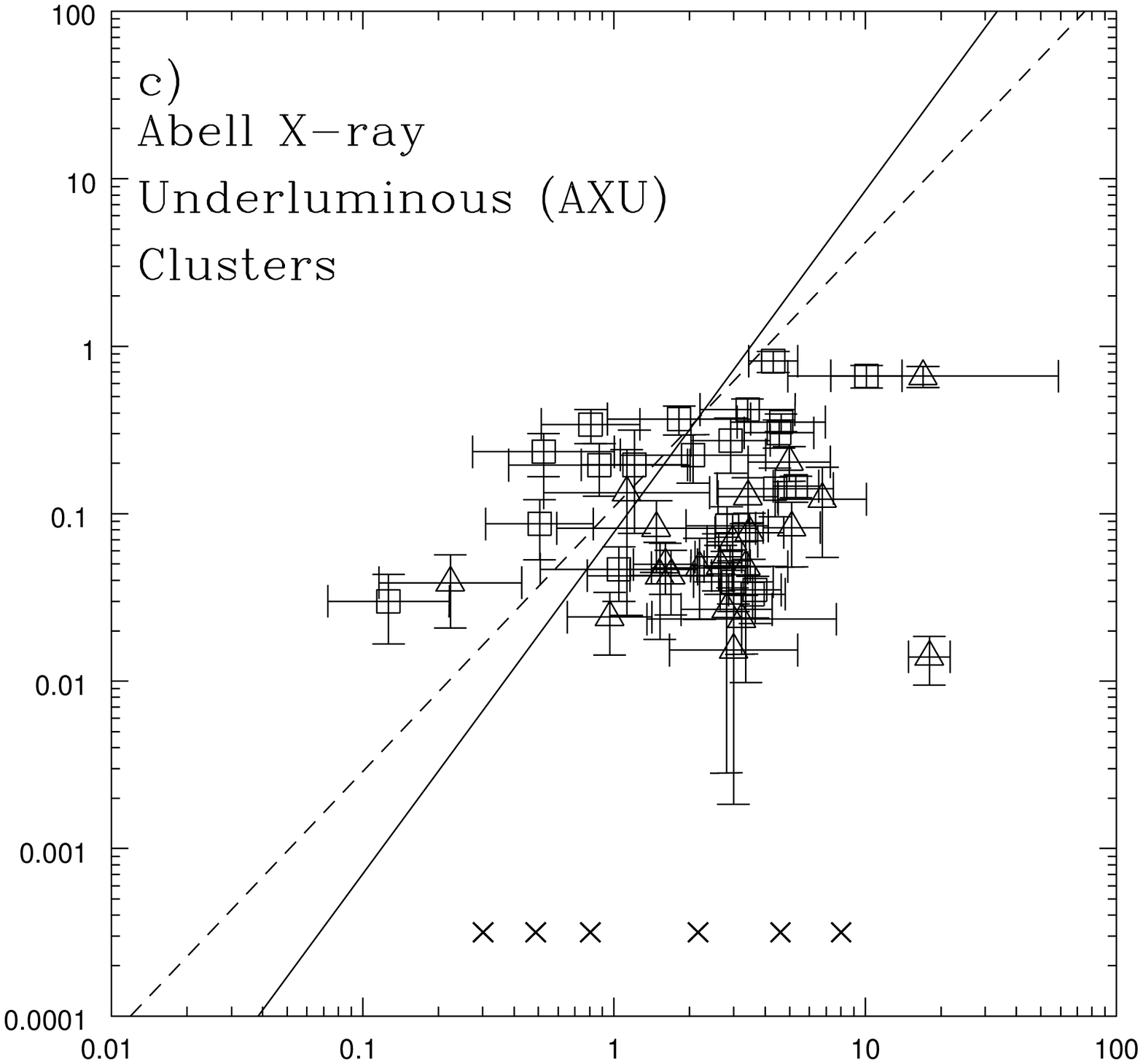}}
\end{minipage}
\begin{minipage}{0.36\textwidth}
\resizebox{\hsize}{!}{\includegraphics{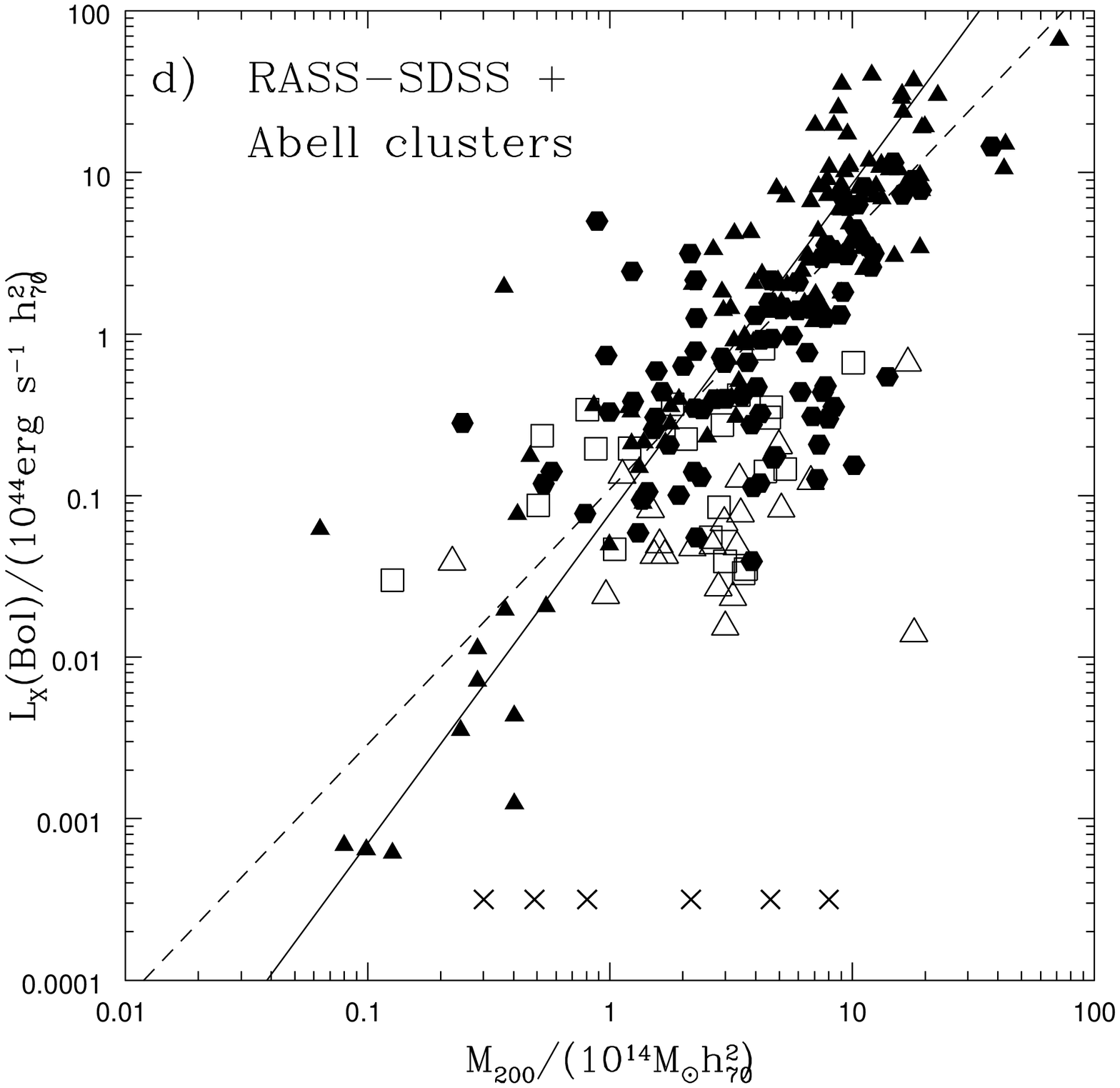}}
\end{minipage}
\begin{minipage}{0.36\textwidth}
\resizebox{\hsize}{!}{\includegraphics{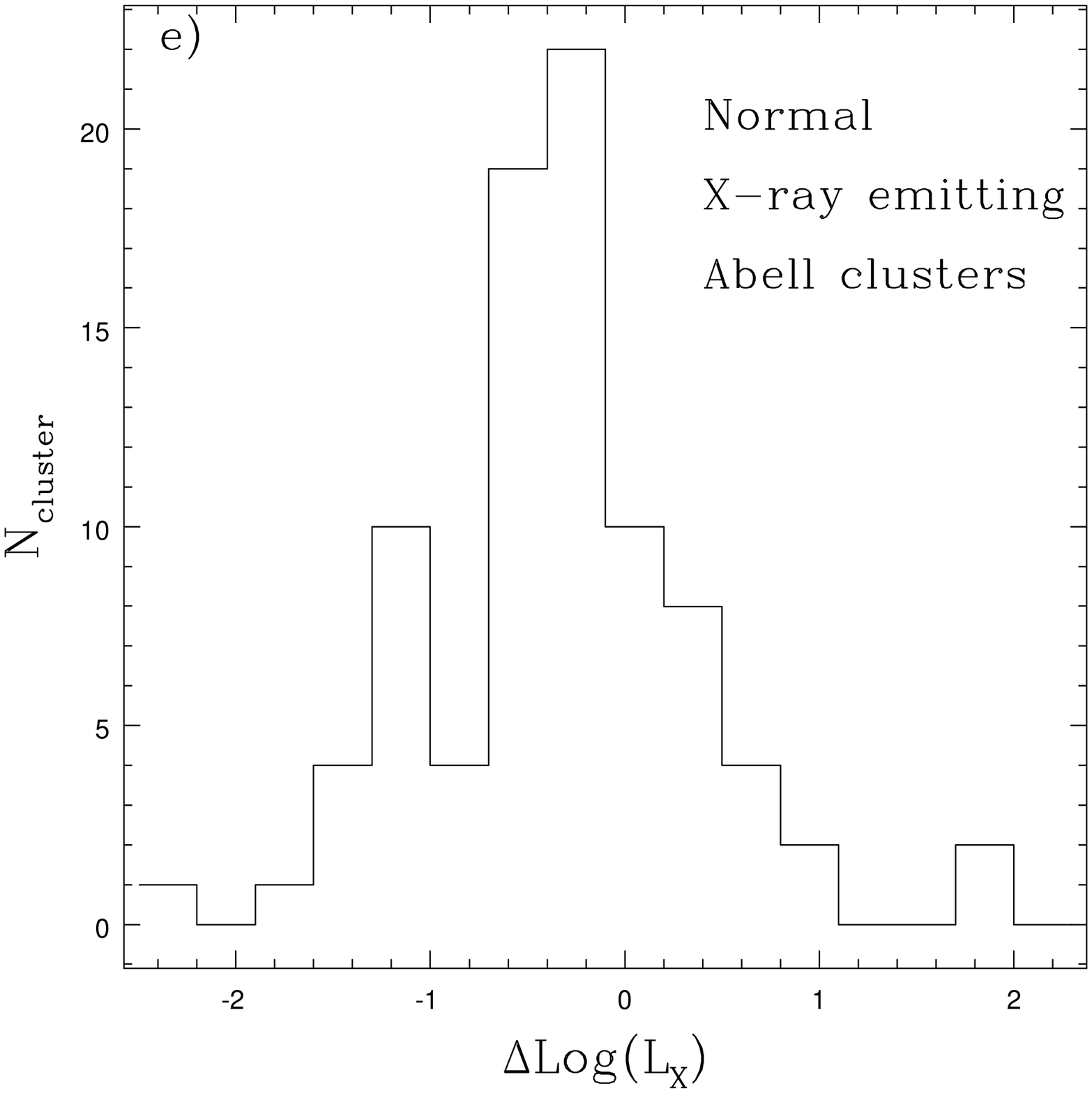}}
\end{minipage}
\begin{minipage}{0.36\textwidth}
\resizebox{\hsize}{!}{\includegraphics{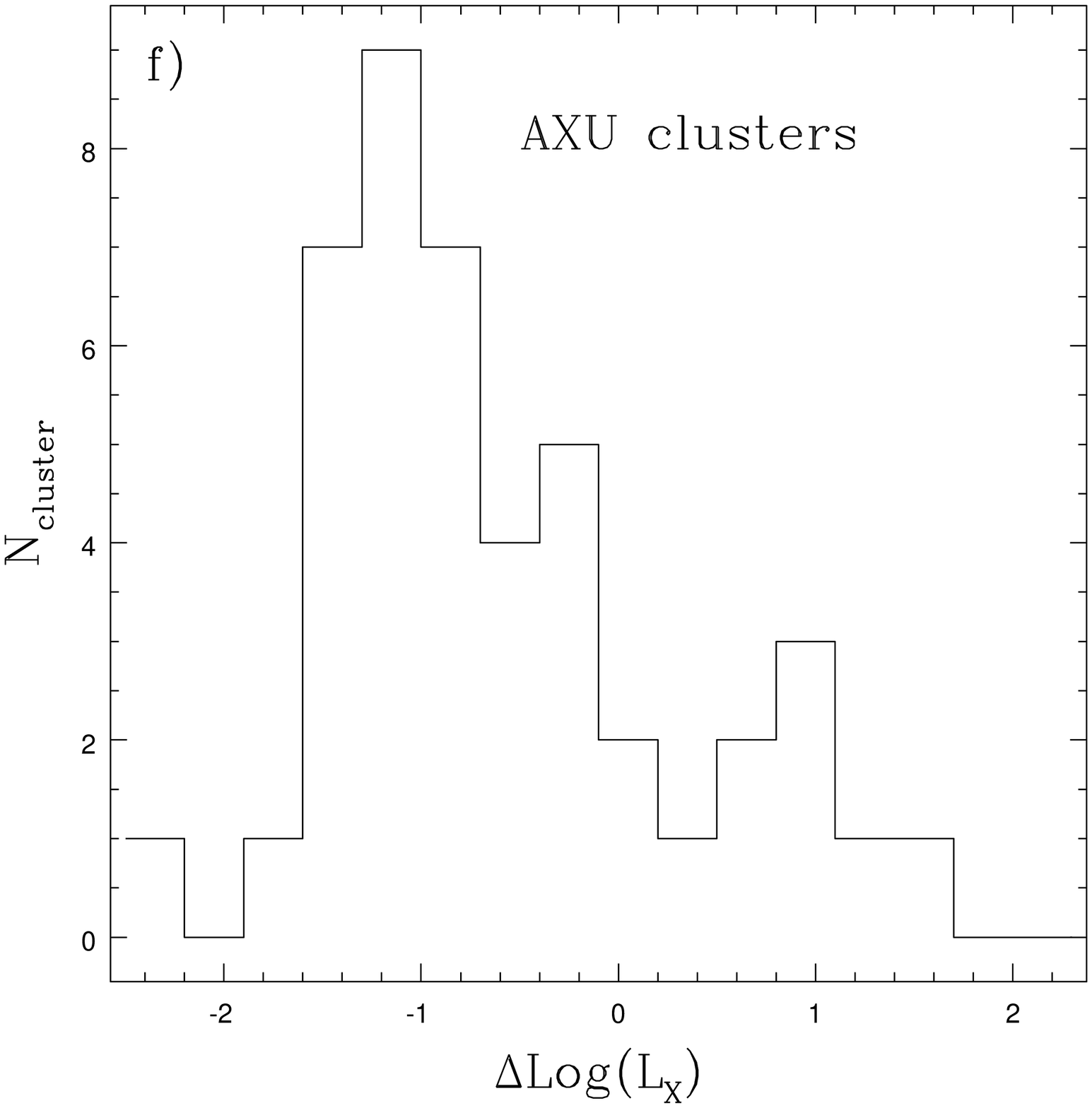}}
\end{minipage}
\end{center}
\caption{
$L_X-M_{200}$ relation. Panel $a)$ shows the $L_X-M_{200}$ of the
X-ray selected RASS-SDSS cluster sample (filled triangles). Panel $b)$
shows the location of the Normal X-ray emitting Abell clusters (filled
dots) relatively to the best fit obtained in the X-ray selected
sample. Panel $c)$ shows the location of the AXU systems in the same
diagram. The empty squares are the Abell clusters with marginally
significant X-ray emission and the empty triangles are the Abell
clusters without X-ray emission (upper limits, see text for the
explanation), the crosses are the Abell clusters for which the GCA
method was not able to calculate the $L_X$ upper limit (they are all
plotted at $L_X=10^{-40.5}$ erg$s^{-1}$). Panel $d)$ shows the
$L_X-M_{200}$ relation for the RASS-SDSS plus the whole Abell
sample. The symbols in the panel have the same meaning as in the
previous three panels. The solid line in the 4 panels is the best fit
line obtained in the whole X-ray selected sample of panel $a)$ and the
dashed line is the best fit obtained from the subsample of 69
RASS-SDSS clusters for which the mass is calculated as for the Abell
clusters. Panel $e)$ shows the distribution of the residuals of the
normal X-ray emitting clusters. Panel $f)$ shows the same distribution
for the AXU clusters. The residuals are defined as $\Delta
log(L_X)=log(L_{X,m})-log(L_{X,p})$, where $L_{X,m}$ is the measured
cluster X-ray luminosity and $L_{X,p}$ is the $L_X$ predicted by the
$L_X-M_{200}$ X-ray relation. }
\label{LX}
\end{figure*}

Cluster members are selected among SDSS galaxies with available
redshifts, as follows.  First, we select only galaxies within a circle
of 2.15 Mpc radius (the Abell radius). We then group together those
galaxies with intergalaxy velocity differences less than a critical
value that depends on the total number of galaxies along the
line-of-sight, according to the relation adopted by Adami et
al. (1998a).  This allows us to define the cluster limits in velocity
space. As an additional step, we apply the membership selection
algorithm of Katgert et al. (2004) to {\em all} the galaxies (also
outside an Abell radius) with velocities within the limits defined
with the gapper procedure.  This algorithm takes into account both the
velocities and the clustercentric positions of the galaxies. The
method is identical to that of den Hartog \& Katgert (1996) when the
cluster sample contains at least 45 galaxies, and it is a simplified
version of it for smaller samples (for more details, see Appendix A in
Katgert et al. 2004). It requires a cluster centre to be defined. When
possible, we adopt the X-ray centre for this.  However some clusters
do not have secure X-ray detection, in which case the X-ray centre
cannot be accurately defined. In those cases we take the position of
the brightest cluster member as the cluster centre (see, e.g., Biviano
et al. 1997). In any case, the analysis of clusters identified in
cosmological numerical simulations indicate that the choice of the
centre is not critical for a correct performance of the membership
selection algorithm (Biviano et al. in preparation).

Only Abell clusters with at least 10 galaxy members are selected,
since 10 is the minimum number of cluster members in order to
calculate in a reasonable way the cluster mass and velocity dispersion
(Girardi et al. 1993). Among the 280 Abell clusters in the region
covered by DR3, 179 fulfil this requirement. Among these clusters, 38
are affected by problems of contamination, due to the presence of a
close companion or a second system along the same line-of-sight but at
different redshift and 4 show large discrepancies between our estimate
of $z_c$ and the value derived from the literature or the $z_c-m_{10}$
relation (Postman 1985). Those systems are excluded from our final
sample. Hence we are left with a sample of 138 Abell clusters, listed
in the Appendix, along with their global properties. As shown in
Fig. \ref{z_dist}, the considered cluster sample comprises only nearby
systems ($ z < 0.25$) at the mean redshift of 0.1. As the X-ray
reference sample, the optically selected cluster sample covers the
entire range of masses and X-ray/optical luminosities, from the
low-mass ( faint X-ray/optical luminosity) regime ($2\times 10^{13}
M{\odot}$) to the high-mass (high X-ray/optical luminosity) regime
($3\times 10^{15} M{\odot}$). We point out that the two cluster
samples (X-ray and optically selected) considered in this work are not
complete. However, for the purpose of this work we do not need
complete cluster samples but clean X-ray and optically selected
cluster samples spanning the all cluster mass and luminosity
range. The X-ray and optically selected cluster samples used in this
work fulfill these requirements.

\subsection{Optical luminosities}
The estimate of the optical luminosity of a cluster, $L_{op}$,
requires subtraction of the foreground and background galaxy
contamination. We consider two different approaches to the statistical
subtraction of the galaxy background. We compute the local background
number counts in an annulus around the cluster and a global background
number counts from the mean of the magnitude number counts determined
in five different SDSS sky regions, each with an area of 30
$\rm{deg^2}$. In our analysis we show the results obtained using the
optical luminosity estimated with the second method, since the two
methods produces only marginal differences in the $L_{op}$ estimates.
The cluster magnitude number counts in the virial region are
obtained by subtracting from the galaxy counts measured within
$r_{200}$ the local (global) field counts rescaled to the cluster
area. The cluster magnitude number counts are converted in luminosity
number counts after dereddening, K-correcting and transforming in
absolute magnitudes the apparent magnitudes. The cluster Optical
Luminosity is then obtained simply by summing up the luminosity number
counts multiplied by the mean luminosity of the bin. The reader is
referred to paper I of this series for the details about the
comparison between optical luminosities obtained with different
background subtraction methods and for the other technical details.

\subsection{Velocity dispersions and virial masses}
The virial analysis (see, e.g., Girardi et al. 1998) is performed on
the clusters with at least 10 member galaxies. The velocity dispersion
is computed on the cluster members, using the biweight estimator
(Beers et al. 1990). The virial masses are corrected for the surface
pressure term (The \& White 1986) by adopting a profile of Navarro et
al.  (1996, 1997; NFW hereafter) with a concentration parameter, $c$,
that depends on the initial estimate of the cluster virial mass
itself.  The $c$--mass relation is given by $c=4 \times
(M/M_{KBM})^{-0.102}$ where the slope of the relation is taken from
Dolag et al. (2004), and the normalization $M_{KBM} \simeq 2 \times
10^{15} M_{\odot}$ from Katgert et al. (2004). The clusters in our
sample span a range $c \simeq 3$--6.

Correction for the surface pressure term requires knowledge of the
$r_{200}$ radius, for which we adopt Carlberg et al.'s (1997)
definition (see eq.(8) in that paper) as a first guess. After the
virial mass is corrected for the surface pressure term, we refine our
$r_{200}$ estimate using the virial mass density itself. Say $M_{vir}$
the virial mass (corrected for the surface term) contained in a volume
of radius equal to a chosen observational aperture, $r_{ap}$, that we
have set equal to the Abell radius, 2.15 Mpc. The radius $r_{200}$ is
then given by:
\begin{equation}
r_{200} \equiv r_{ap} \, [\rho_{vir}/(200 \rho_c)]^{1/2.4}
\label{e-r200}
\end{equation}
where $\rho_{vir} \equiv 3 M_{vir}/(4 \pi r_{ap}^3)$ and $\rho_c(z)$
is the critical density at redshift $z$ in the adopted cosmology. The
exponent in eq.(\ref{e-r200}) is the one that describes the average
cluster mass density profile near $r_{200}$, as estimated by Katgert
et al. (2004) for an ensemble of 59 rich clusters. 

For consistency the $c$--mass relation is used to interpolate (or, in
a few cases, extrapolate) the virial mass $M_{vir}$ from $r_{ap}$ to
$r_{200}$, yielding $M_{200}$. From $M_{200}$ the final estimate of
$r_{200}$ is obtained, using the definition of $M_{200}$ itself.

Even if the completeness level of the SDSS spectroscopic sample
is very high, in the central regions of galaxy clusters such a level
is likely to drop because fibers cannot be placed closer than 55
arcsec. We estimate that the spectroscopic completeness drops to $\sim
70$\% in the central $\sim 0.1$ Mpc cluster regions. This affects the
observed number density profile of a cluster, and hence our virial
mass estimates (see, e.g., Beers et al. 1984). Using the average
cluster number density profile we estimate that this effect of
incompleteness translates into an average over-estimate of the virial
mass of only $\sim 5$\% (see Paper III of the series for more details
about this estimate). Since the effect is very small, and much smaller
than the observational uncertainties, we neglect this correcting
factor in the following analysis.

\subsection{X-ray luminosities}
In order to create a homogeneous catalog of X-ray cluster properties, we
search for the X-ray counterparts of all the 137 Abell clusters,
and compute their X-ray luminosity, $L_X$, using
only RASS data.

X-ray luminosities are calculated with the growth curve
analysis (GCA) method used for the NORAS and REFLEX cluster surveys
(B\"ohringer et al. 2000) based on the RASS3 data base (Voges et
al. 1999). The GCA method is optimized for the detection of the
extended emission of clusters by assessing the plateau of the
background subtracted cumulative count rate curve. We use as a final
result the total flux inside the radius $r_{200}$ which is corrected
for the missing flux estimated via the assumption of a standard
$\beta$-model for the X-ray surface brightness (see B\"ohringer et
al. 2000 for more details). The correction is typically only $8 -
10\%$ illustrating the high effectiveness of the GCA method to sample
the flux of extended sources.

We check by eye all the X-ray sources associated to the Abell
clusters. We find a secure X-ray detection for 86 systems out of the
137 isolated and well classified Abell clusters. Other 27 have a
marginally significant detection (between 2 and 3 $\sigma$) and
other 24 do not have clear X-ray emission (detection level $\sim
1 \sigma$ or no detection at all). The GCA method provides an
estimate of the X-ray detection also in case of dubious X-ray
detection, but the percentage error is higher than 80\% and the
estimate has to be considered as an upper limit. In 7 cases out of the
24 systems withot X-ray detection the GCA method fails completely to
provide an estimate of $L_X$. The X-ray luminosity ended up to be
negative after the background subtraction. For those systems the X-ray
luminosity is set equal to zero. We will discuss in detail the nature
of these 27+24 clusters with marginal or no X-ray detection in the
following sections. We refer to these 51 sysstems in the next
paragraph as ``clusters without secure X-ray detetction''.

\begin{table*}
\begin{center}
\begin{tabular}[b]{cc|c|ccccc}
\hline
\multicolumn{2}{c|}{A-B relation}& \multicolumn{1}{c|}{sample}& \multicolumn{5}{c|}{}\\ \hline
\renewcommand{\arraystretch}{0.2}\renewcommand{\tabcolsep}{0.05cm}
A& B& &$\alpha $ & $\beta$ & $\sigma$ &$\sigma _B$ & $\sigma_A$ \\ \hline
$L_X(Bol)$ & $M_{200}$ & X-ray   &2.04 $\pm$        0.08&        -1.11  $\pm$       0.06&         0.16  &       0.21   &      0.43 \\
	&			& Abell   &2.19 $\pm$        0.14&        -1.67  $\pm$       0.14&         0.23  &       0.32   &      0.48 \\
      &                & A+X-ray &2.12 $\pm$        0.08&        -1.32  $\pm$       0.07&         0.22  &       0.29   &      0.48 \\
      &                & $A(P_{DS}>0.1)+X-ray$ &2.06 $\pm$        0.08&        -1.21  $\pm$       0.06&         0.18  &       0.23   &      0.46 \\
\hline
$L_X(Bol)$ & $L_{op}$  & X-ray &1.72 $\pm$          0.08&        -0.98  $\pm$       0.07&         0.17  &       0.19   &      0.31 \\
	&			& Abell &2.01 $\pm$          0.15&        -1.17  $\pm$       0.09&         0.20  &       0.28   &      0.35 \\
      &                & A+X-ray &1.87 $\pm$        0.08&        -1.08  $\pm$       0.06&         0.19  &       0.25   &      0.35 \\
\hline
$L_{op}$ & $M_{200}$  & X-ray    &0.88 $\pm$        0.03&        -0.08  $\pm$       0.02&         0.13  &       0.18   &      0.16 \\
	&			& Abell   &0.80 $\pm$        0.07&        -0.01  $\pm$       0.04&         0.14  &       0.21   &      0.22 \\
      &               & A+X-ray  &0.83 $\pm$        0.03&        -0.05  $\pm$       0.03&         0.14  &       0.20   &      0.19 \\
\hline
\end{tabular}
\caption{ The table lists the best fit parameters for the relations
between several global cluster quantities, i.e. the bolometric X-ray
luminosity, $L_X(Bol)$, the virial mass, $M_{200}$, and the $i$-band
optical luminosity $L_{op}$, for different samples of galaxy clusters.
The 'X-ray' refers to the X-ray selected systems with known mass,
taken from the RASS-SDSS galaxy cluster catalog (Paper III).  The
'Abell' refers to the whole Abell sample considered in this work. The
'A+X-ray' refers to the Abell sample plus the X-ray selected cluster
sample. The '$A(P_{DS}>0.1)+X-\rm{ray}$' refers to the X-ray selected
clusters plus the Abell sample without the clusters with unsecure X-ray
detection and the systems with high level of subclustering. The table
lists three estimations of the scatter for each relation: $\sigma$ is
the orthogonal scatter of the A-B relation (where $A=10^{\beta} \times
B^{\alpha}$), $\sigma_A$ is the scatter in the A variable and
$\sigma_B$ is the scatter in the B variable. All the scatter values in
the table are expressed in dex, while all the errors are given at the
95\% confidence level.}
\label{table1}
\end{center}							   
\end{table*}

\begin{figure}
\begin{center}
\begin{minipage}{0.5\textwidth}
\resizebox{\hsize}{!}{\includegraphics{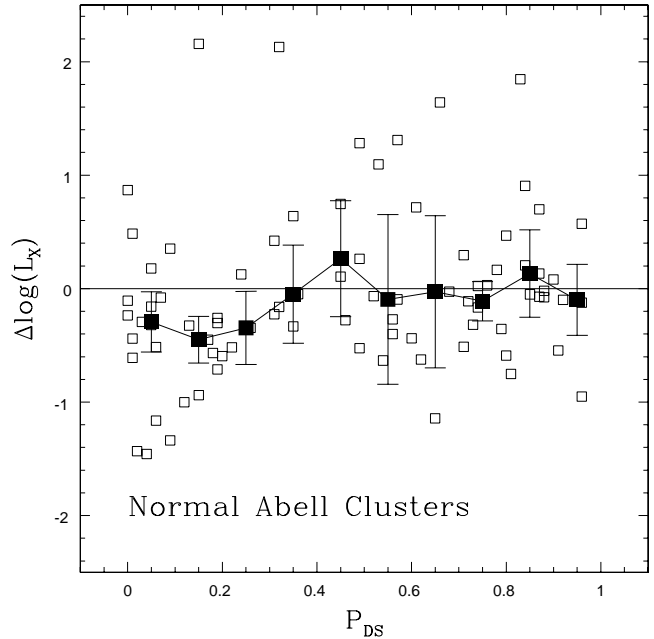}}
\end{minipage}
\end{center}
\caption{The X-ray luminosity residuals $\Delta log(L_X)$ 
from the best-fit $L_X-M_{200}$ relation of normal Abell clusters,
vs. the Dressler \& Shectman parameter $P_{DS}$. Systems with
$P_{DS}<0.1$ are considered to be characterized by
subclustering. Filled squares with error bars represent the mean and
dispersion of all points in bins of $P_{DS}$.}
\label{dlog_lx1}
\end{figure}

\begin{figure}
\begin{center}
\begin{minipage}{0.5\textwidth}
\resizebox{\hsize}{!}{\includegraphics{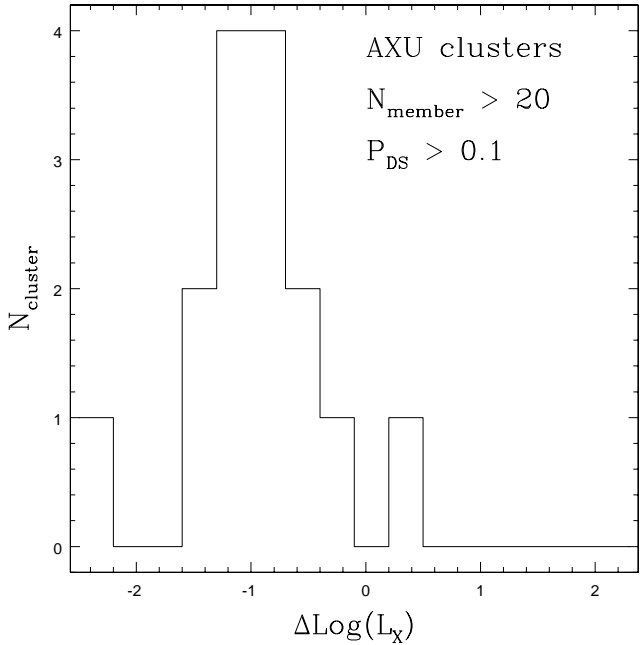}}
\end{minipage}
\end{center}
\caption{
Distribution of the residuals along the $log(L_X)$ axis for the Abell
clusters without secure X-ray detection with more than 20
spectroscopic members within 1 abell radius and with $P_{DS} >
0.1$. The residuals are defined as $\Delta
log(L_X)=log(L_{X,m})-log(L_{X,p})$, where $L_{X,m}$ is the measured
cluster X-ray luminosity and $L_{X,p}$ is the $L_X$ predicted by the
$L_X-M_{200}$ X-ray relation.}
\label{n_20}
\end{figure}

\begin{figure*}
\begin{center}
\begin{minipage}{0.48\textwidth}
\resizebox{\hsize}{!}{\includegraphics{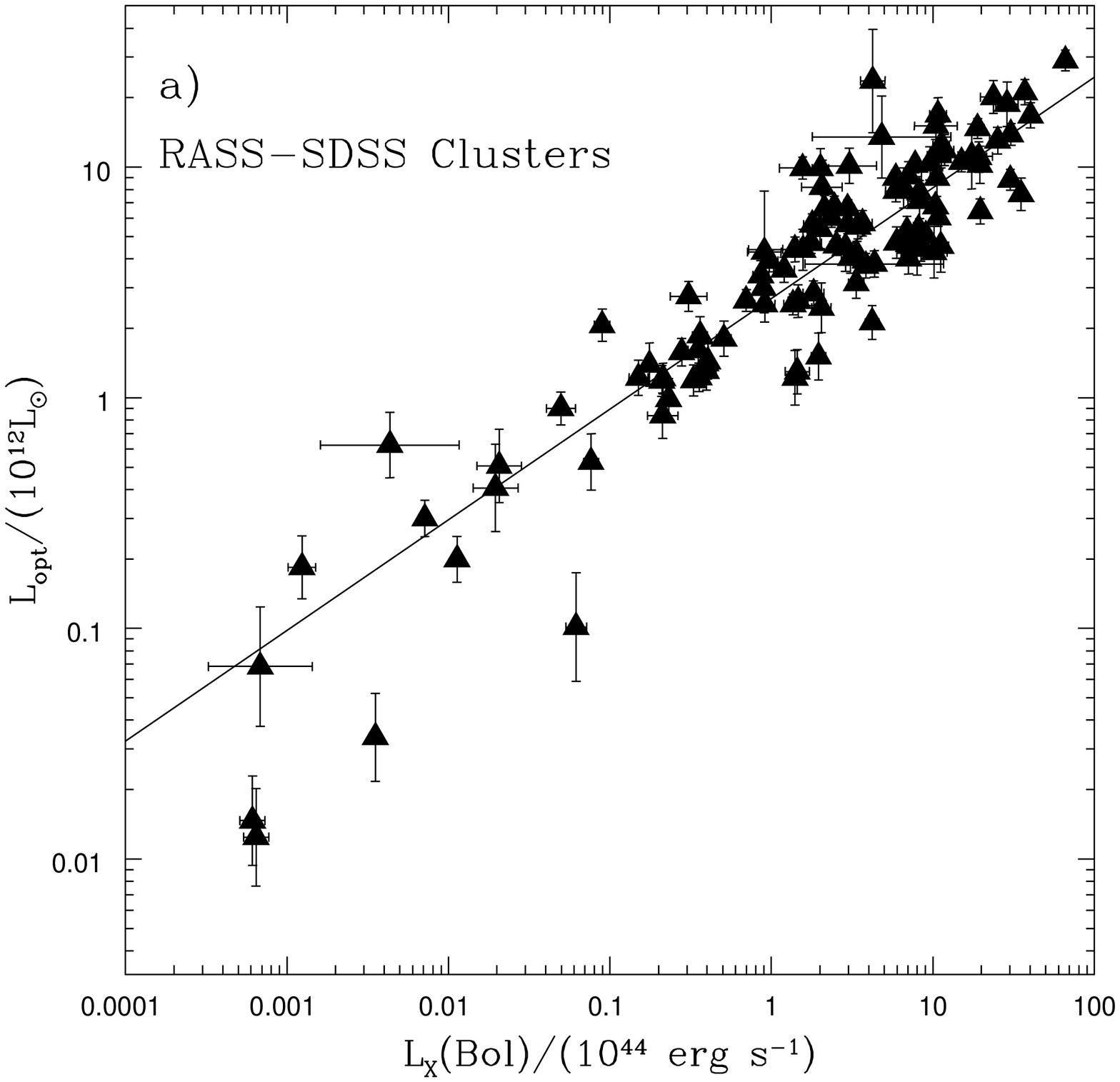}}
\end{minipage}
\begin{minipage}{0.48\textwidth}
\resizebox{\hsize}{!}{\includegraphics{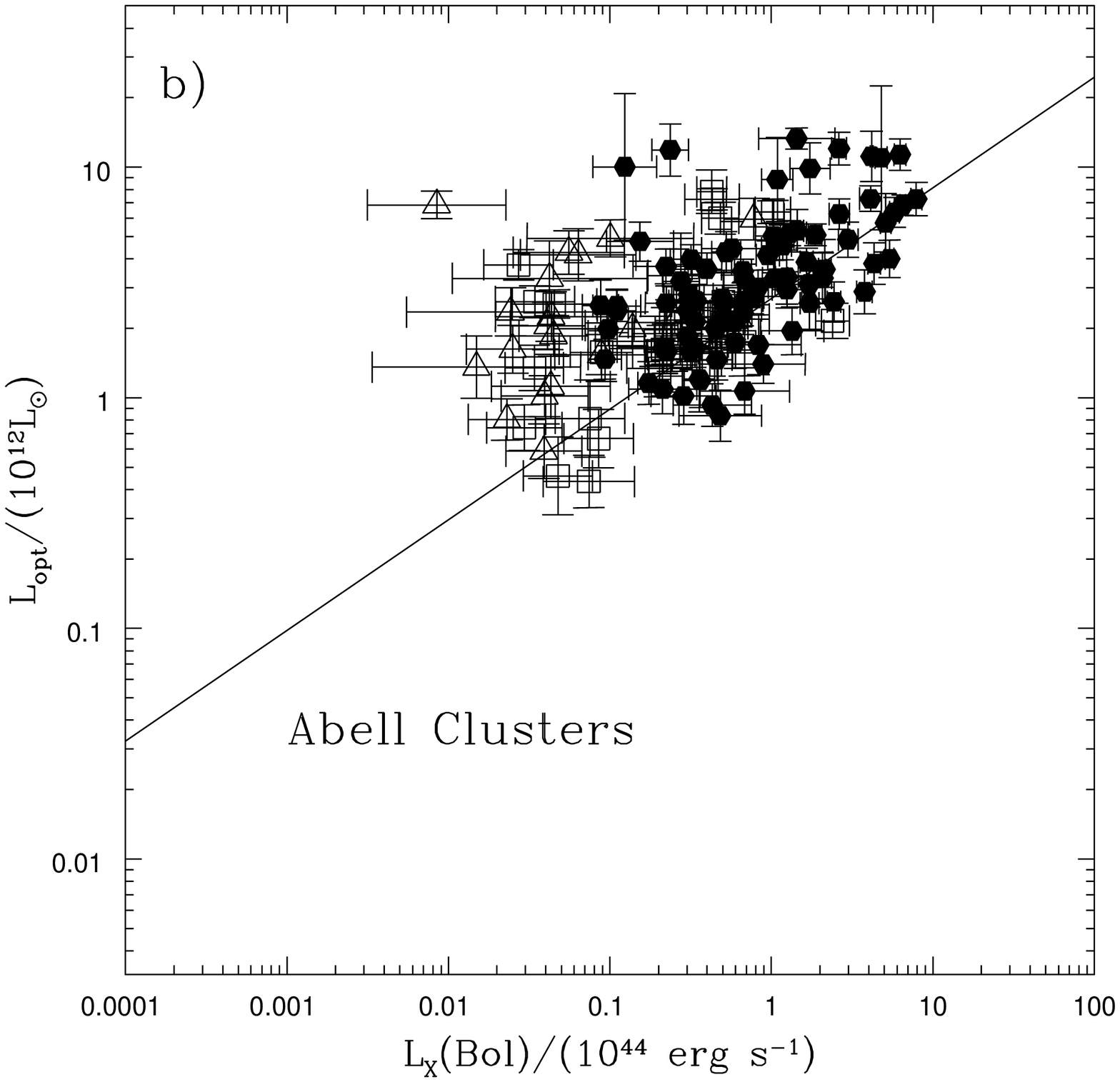}}
\end{minipage}
\end{center}
\caption{
$L_X-L_{op}$ relation. in panel $a)$ the filled triangles are the
X-ray selected clusters of the RASS-SDSS sample of
Paper III. In panel $b)$ the filled points are the normal X-ray
emitting Abell clusters, the empty triangles are the AXU clusters
with a marginally significant X-ray detection, the empty squares are
the AXU clusters with no detection. The solid line is the best fit
obtained from the RASS-SDSS clusters. The optical luminosity is
computed in the $i$-band.}
\label{lxlo}
\end{figure*}

\section{X-ray versus optical properties}
In this section we analyze the relations among the Bolometric X-ray
luminosity, the cluster mass, $M_{200}$, and the optical $i$-band
luminosity, $L_{op}$. The Bolometric X-ray luminosity is derived by
correcting the X-ray luminosity in the ROSAT energy band (0.1-2.4 keV)
with the bolometric correction corresponding to the cluster
temperature. The cluster temperature is estimated from the cluster
mass using the $T_X-M_{200}$ relation given in Paper III. We perform
an orthogonal linear regression in logarithmic space for each of the
analyzed relations. The orthogonal regression is performed with the
software package ODRPACK (Akritas $\&$ Bershady 1996). Table
\ref{table1} lists the values of the best fit parameters and the
scatter for all the analysed correlations.

\subsection{The $M-L_X$ relation and the Abell X-ray Underluminous Clusters}
Panel $a)$ of Fig.\ref{LX} shows the $L_X-M_{200}$ relation obtained
from the X-ray selected RASS-SDSS galaxy cluster sample. The RASS-SDSS
galaxy cluster sample comprises 102 systems. For 69 of them the mass,
$M_{200}$, is calculated through the dynamical analysis as explained
in section 2.3. For the remaining 33 objects the mass is calculated
using the known ICM temperature in the M-T relation given in paper
III. The solid line in the panel $a)$ of Fig.\ref{LX} shows the best
fit line obtained with the whole sample (102 clusters) and the dashed
line shows the best fit line obtained using the 69 clusters for which
the mass is calculated as for the Abell Clusters.

Panel $b)$ of Fig.\ref{LX} shows the location of the 86 Abell clusters
with clear X-ray detection in RASS relatively to the best fit obtained
on the X-ray selected sample. Panel $c)$ shows the behaviour of the
Abell systems without secure X-ray detection in RASS in the same
diagram.

As shown by panel $d)$ of Fig. \ref{LX}, the scatter of the
$L_X-M_{200}$ relation increases significantly when the Abell clusters
are added to the sample of RASS-SDSS clusters. The best fit parameters
of the $L_X-M_{200}$ relation obtained by considering the Abell and
RASS-SDSS clusters together is consistent with the relation found for
the RASS-SDSS clusters only (see Table \ref{table1}). However, the
orthogonal scatter increases from 44 to 65\%. The RASS-SDSS
clusters sample comprises several cluster (10 objects) at redshift
higher than the redshift range of the Abell clusters. Thus, to check
the possible effect of evolution on the scatter of the considered
relation, we perform the analysis considering the RASS-SDSS clusters
in the same redshift range as the Abell clusters. The resulting
correlations are perfectely consistent with the results listed in
Table \ref{table1} for all the considered cluster sample. The
scatter increase is not only due to the Abell clusters without clear
X-ray detection. Instead, a large contribution to the increase of the
scatter is given by the normal Abell clusters which show a high level
of subclustering. In fact, the presence of substructures causes the
cluster mass to be overestimated. Therefore the systems presenting
subclustering should deviate from the relation.  We quantify the
presence of galaxy substructures in the whole Abell cluster sample
through the Dressler \& Shectmam (1988) statistical test. This test
looks for deviations of the local velocity mean and dispersion from
the global values. Here we adopt the slightly modified version of the
test introduced by Biviano et al. (2002). We call $P_{DS}$ the
probability that a cluster does {\em not} contain substructures
according to the Dressler \& Shectman test.  We find that the fraction
of clusters with a probability $>0.90$ ($P_{DS} < 0.1$) of having
significant substructure is somewhat low, 20\%, compared to the
results of previous studies (e.g. Dressler \& Shectman 1988; Biviano
et al. 1997). This is not surprising. We remind the reader that the
137 Abell clusters in our sample are selected to be relatively
isolated and free of major contaminations along the line-of-sight (see
section
\ref{s-sample}). Anyhow, as shown in Fig. \ref{dlog_lx1} the cluster
with values of $P_{DS}$ lower than 0.1 have the largest negative
residuals from the best fit line.  When the 20\% of clusters with high
level of subclustering (together with the Abell systems with unsecure
X-ray detection) are excluded from the linear regression, the best fit
parameters and the scatter of the relation are consistent with the
values found in the case of the RASS-SDSS cluster sample.  Table
\ref{table1} lists the results of this linear regression in the line
corresponding to the $A(P_{DS}>0.1)+X-\rm{ray}$ sample, which refers to the
Abell clusters with $P_{DS} > 0.1$ plus the RASS-SDSS systems.

In order to carachterize the different behavior of the normal Abell
Clusters and the Abell systems without secure X-ray detection, we
analyse the distribution of the residuals of the Abell clusters
relatively to the RASS-SDSS  $L_X-M_{200}$ relation, along the
$\log(L_X)$ axis. The residuals are defined as $\Delta
\log(L_X)=\log(L_{X,m})-\log(L_{X,p})$, where $L_{X,m}$ is the
measured cluster X-ray luminosity and $L_{X,p}=0.0776 \,
M_{200}^{2.04}$ is the $L_X$ predicted by the $L_X-M_{200}$ relation
(see Table \ref{table1}).  Hence, a negative value of the residual
indicates that the cluster has a low X-ray luminosity for its mass.

Panel $e)$ of Fig.\ref{LX} shows the distribution of the residuals of
the normal Abell clusters. The median of the distribution is at $-0.3
\pm 0.3$ and it moves to $-0.1 \pm 0.3$ when the clusters with high
level of subclustering are excluded. This confirms that those systems
obey the same $L_X-M_{200}$ relation as the RASS-SDSS clusters.

Panel $f)$ of Fig.\ref{LX} shows the same distribution for the Abell
clusters without secure X-ray detection (except clusters with zero
$L_X$).  The median of the distribution is at $-0.9 \pm 0.4$, which
gives an indication that those clusters are {\em not} on the same
$L_X-M_{200}$ relation. 70\% of those systems have an X-ray luminosity
which is more than 3 times lower than what expected at their mass and
50\% of them have $L_X$ one order of magnitude lower than the
expectation. Hence, the Abell Clusters without secure X-ray detection
appear to be clearly X-ray underluminous for their mass. What causes
this effect? Are those systems real clusters? The poor significance of
the X-ray detection of these systems would suggest that it's a
question of spurious detections in the redshift distribution. That is,
the observed 3D galaxy overdensity of those systems is not due to a
unique massive cluster but to the superposition of two interacting
small groups. In fact, in this case a double peaked velocity
distribution of the two systems could be missclassified as a unique
Gaussian distribution with a large velocity dispersion. As a
consequence the low X-ray luminosity of the two groups would be
associated to the mass of a spurious massive cluster. To check this
possibility we perform several tests. A doubled peak velocity
distribution missclassified as a Gaussian should appear as a
platikurtic distribution (more flat-topped than a Gaussian). This
effect can be quantified with the the robust Tail Index
(T.I. herafter, Beers et al. 1991). Values of the T.I. larger than
unity indicate a leptokurtic distribution (i.e. more centrally peaked
than a Gaussian), while values smaller than unity indicate a
platikurtic distribution. Values close to unity indicate a consintency
with a Gaussian distribution. First, we compute the T.I. values of the
individual cluster velocity distributions, for those clusters with
unsecure X-ray detection with at least 10 member galaxies within
$r_{200}$. 37 out of 51 systems fulfill this requirement. 3 out of 37
have platikurtic distributions and 1 has a leptokurtic one, while all
the remaining distributions are consistent with a Gaussian. The
confidence level used in the test is the 99\%. Therefore, less than
10\% of the clusters are suspected spurious detection. We perform the
same analysis on the normal Abell clusters finding the same percentage
of platikurtic distributions.

As a further test we use the Dressler \& Shectman parameter to
estimate the level of subclustering of those objects. Also this test
is sensitive to the presence of different peaks in the redshift
distribution and could reveal misclassifications. Only 5 clusters out
of 51 systems without secure X-ray detection have values of $P_{DS}$
lower than 0.1 (they comprise the 3 clusters with T.I lower than
1). Hence the fact that a cluster is detected or not in X-ray does not
seem to be related to subclustering in the distribution of cluster
galaxies. 

An aditional cause of uncertainties in the mass estimation is the use
of a small number of spectroscopic members in the measurement. To
check this point, we perform the analysis of the residuals along the
$log(L_X)$ axis for the systems with a high number of
members. Fig. \ref{n_20} shows the distribution of the residuals along
the $log(L_X)$ axis for the Abell clusters without secure X-ray
detection with more than 20 spectroscopic members within 1 Abell
radius and with $P_{DS}>0.1$. The mass estimation of these clusters
with a high number of member galaxies should be less affected by the
systematicis considered so far. The fact that the distribution still
peaks at $-1.0\pm0.3$ confirms that these systems do not lie on the
RASS-SDSS $M-L_X$ relations and that they are on average one order of
magnitude fainter in the X-ray band than what is expected for their
mass. Moreover, in the paper III of this series (Popesso et
al. 2005b) we show that in the case of low level of subclustering the
masses obtained from the dynamical analysis of the cluster members are
consistent with the hydrodynamical mass estimates.

On the basis of these analyses we conclude that the Abell clusters
without secure X-ray detection are not spurious objects and that their
difference with regard to the normal Abell systems and the RASS-SDSS
clusters is physical. Due to their location with regard to the X-ray
$M-L_X$ relation these objects are on average one order of magnitude
fainter than what is expected for their mass. Their marginal detection
or non-detection in X-rays suggests that RASS is too shallow to reveal
the (probably weak) X-ray emission of these systems. Moreover, the
detection depends on parameters not related to the cluster properties
like local RASS exposure, galactic $N_H$, and cluster distance. This
conclusion is supported by the fact that several of these
underluminous X-ray clusters are confirmed to be very faint X-ray
systems by other, independent analyses (Donahue et al. 2002, Ledlow et
al. 2003), based on RASS PSPC pointed observations with longer
exposure times. For these reasons a better physical distinction
between these systems and the normal Abell clusters is the
underluminosity in X-rays of the cluster compared to the RASS-SDSS
relation. However, since the errors on $L_X$ for these clusters is so
large our chosen subdivision is more practical. We call these objects
Abell X-ray Underluminous Clusters (AXU Clusters for short) troughout
the paper.

\subsection{The $L_X-L_{op}$ and the $L_{op}-M$ relations }

Panel $a)$ of Fig. \ref{lxlo} shows the $L_X-L_{op}$ relation for the
RASS-SDSS clusters (the optical luminosity is computed in the
$i$-band).  Panel $b)$ of the same figure shows the $L_X-L_{op}$
relation for the Abell sample. Similarly to what was found for the
$L_X-M_{200}$ relation, the best-fit regression lines obtained using
the RASS-SDSS sample, or the combined RASS-SDSS and Abell samples, are
not significantly different (see Table \ref{table1}).  Also in this
case, the inclusion of the Abell clusters increases the scatter in the
fitted relation. The AXU clusters are the main source of scatter but
also the normal Abell clusters with high level of subclustering
contribute to increase the scatter. The AXU clusters are significantly
offset from the RASS-SDSS $L_X-L_{op}$ relation, while the normal
Abell clusters are not.  The mean residual of the normal Abell
clusters along the $\log(L_{op})$ axis is $0.12 \pm 0.25$, while that
of the AXU clusters is $0.54 \pm 0.20$. Thus, the AXU clusters are
significantly underluminous in X-ray at given optical luminosity
compared to both the normal Abell clusters and the X-ray selected
RASS-SDSS systems.

\begin{figure*}
\begin{center}
\begin{minipage}{0.24\textwidth}
\resizebox{\hsize}{!}{\includegraphics{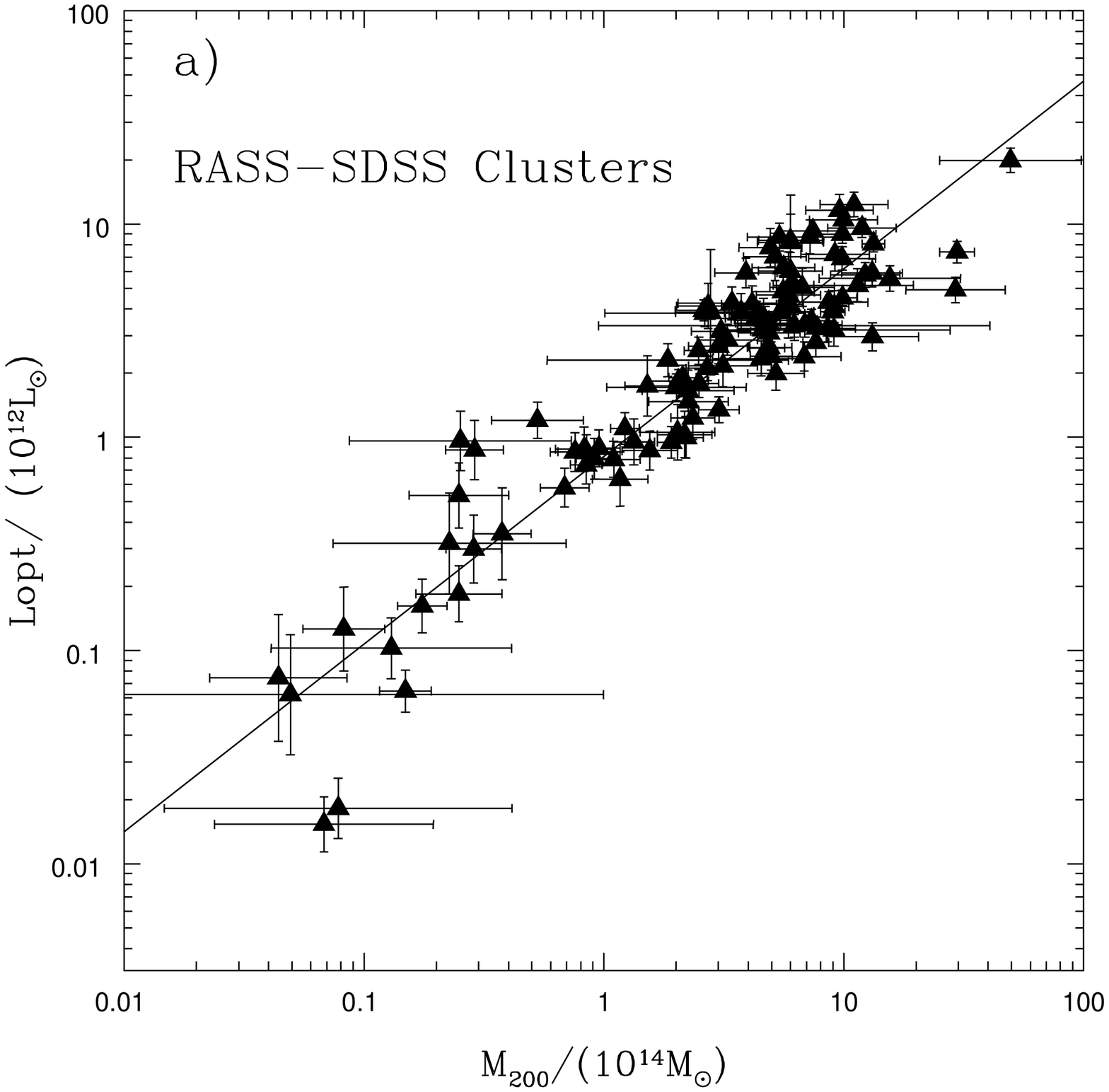}}
\end{minipage}
\begin{minipage}{0.24\textwidth}
\resizebox{\hsize}{!}{\includegraphics{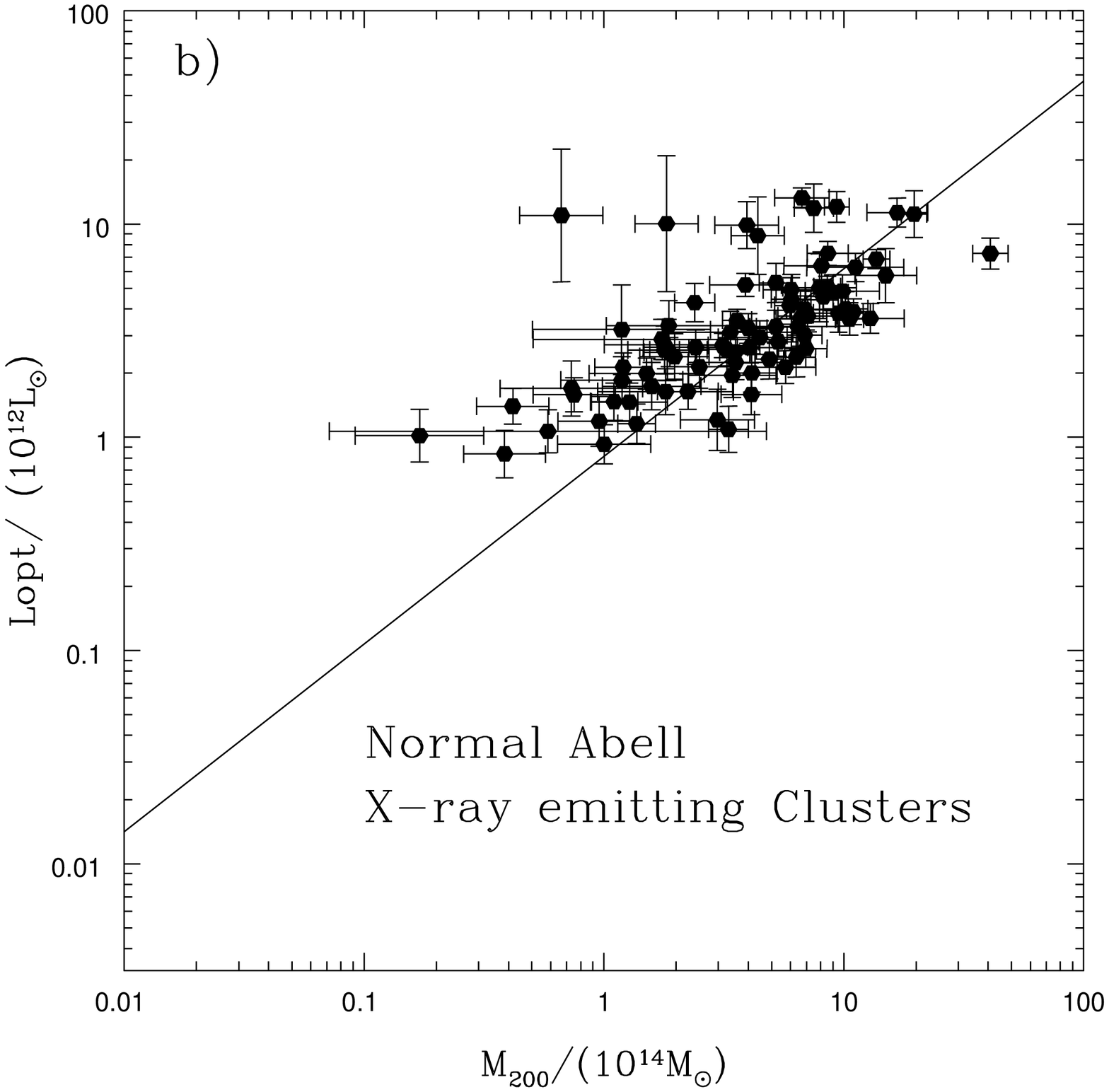}}
\end{minipage}
\begin{minipage}{0.24\textwidth}
\resizebox{\hsize}{!}{\includegraphics{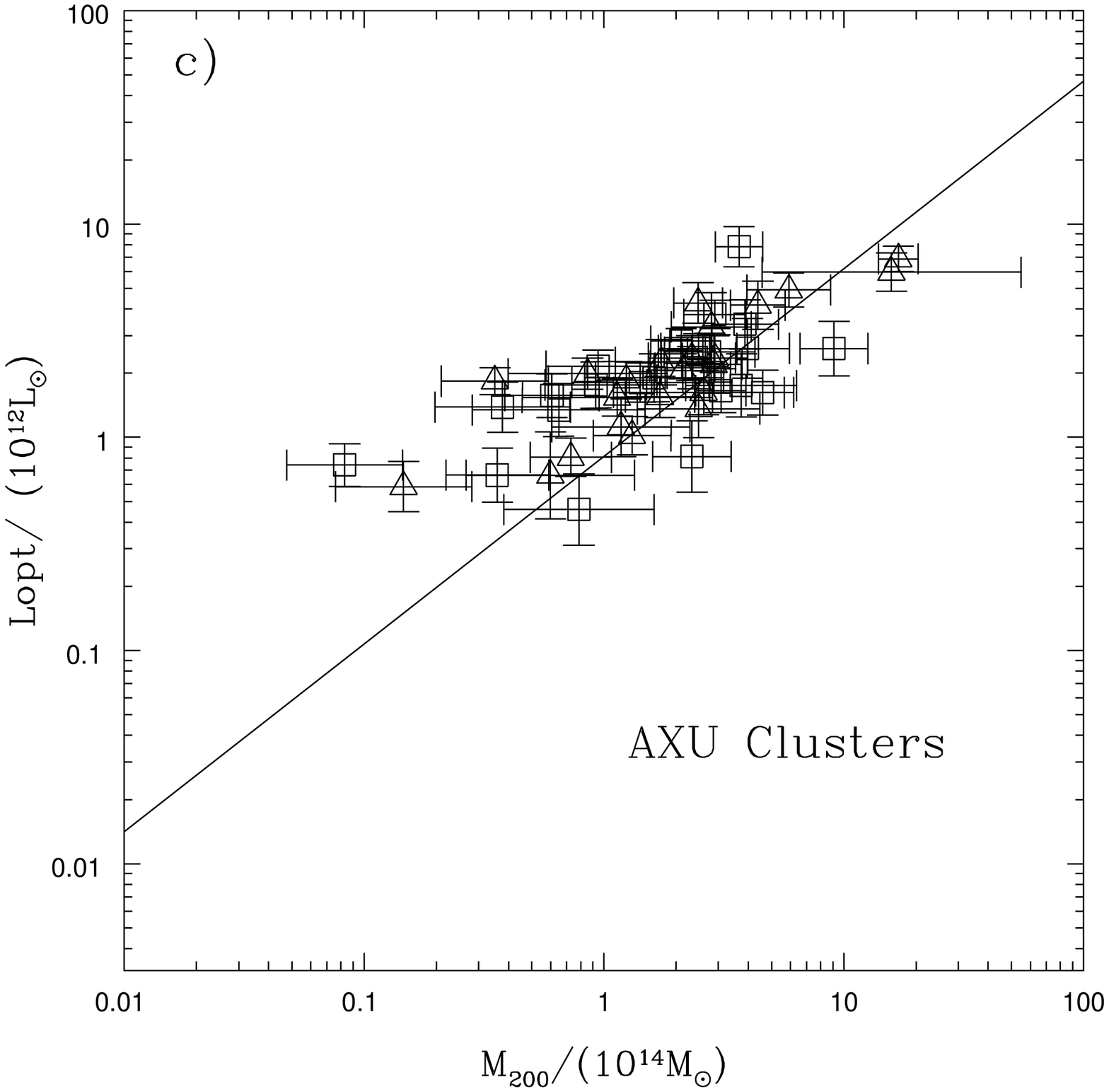}}
\end{minipage}
\begin{minipage}{0.24\textwidth}
\resizebox{\hsize}{!}{\includegraphics{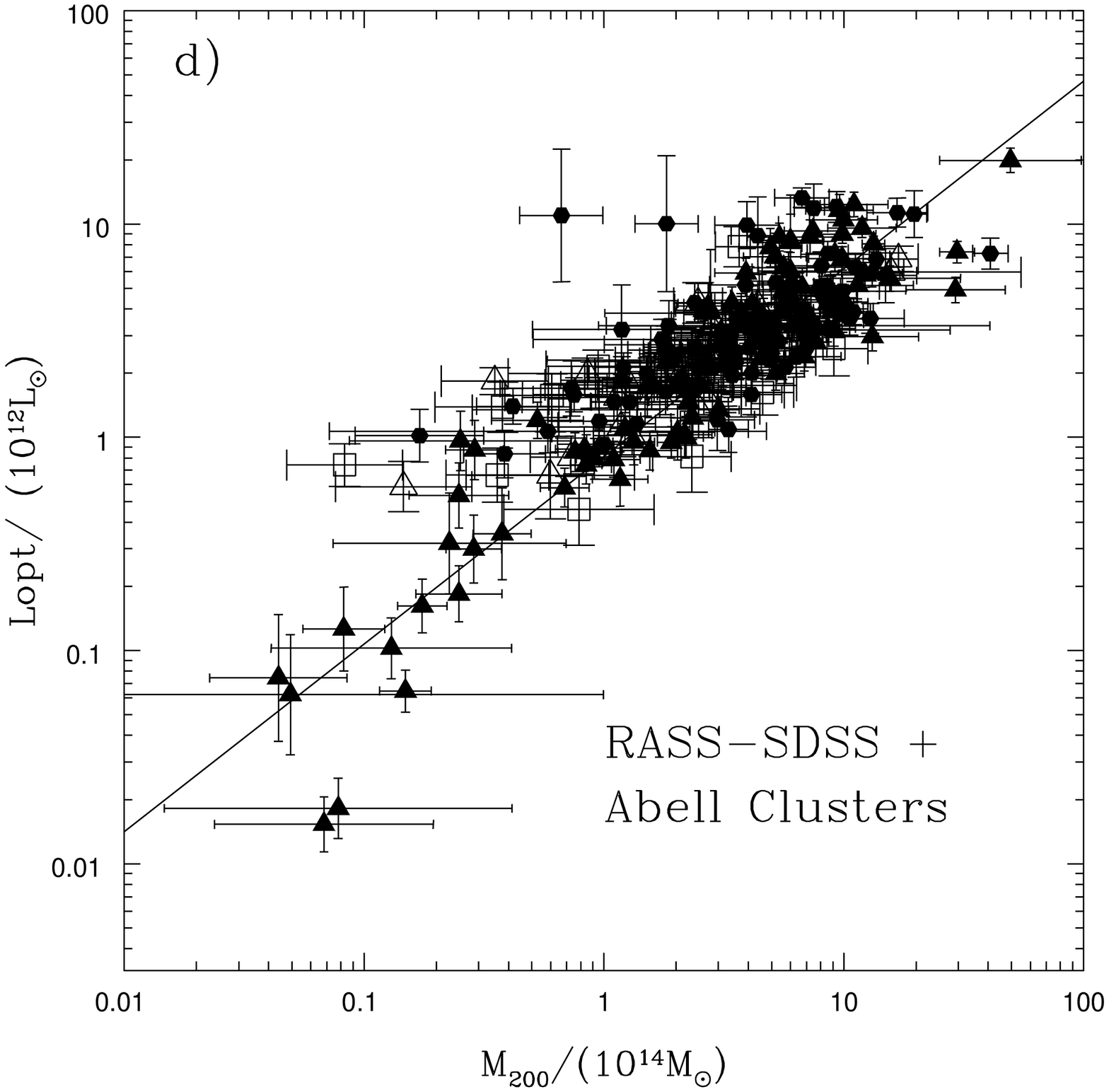}}
\end{minipage}
\end{center}
\caption{ $L_{op}-M_{200}$ relation.  Panel $a)$ shows the
$L_{op}-M_{200}$ of the X-ray selected RASS-SDSS cluster sample
(filled triangles). Panel $b)$ shows the location of the Normal X-ray
emitting Abell Clusters (filled dots) relatively to the best fit
obtained in the X-ray selected sample. Panel $c)$ shows the behaviour
of the AXU systems in the same relation. The empty squares are the AXU
clusters with marginally significant X-ray emission and the empty
triangles are the totally underluminous AXU clusters. Panel $d)$ shows
the $L_{op}-M_{200}$ relation for the RASS-SDSS plus the whole Abell
sample. Symbols in this panel have the same meaning as in the
previous three panels. The solid line in all 4 panels is the best fit
line obtained in the X-ray selected sample of panel $a)$. The optical
luminosity is computed in the $i$-band.}
\label{lom}
\end{figure*}

\begin{figure}
\begin{center}
\begin{minipage}{0.49\textwidth}
\resizebox{\hsize}{!}{\includegraphics{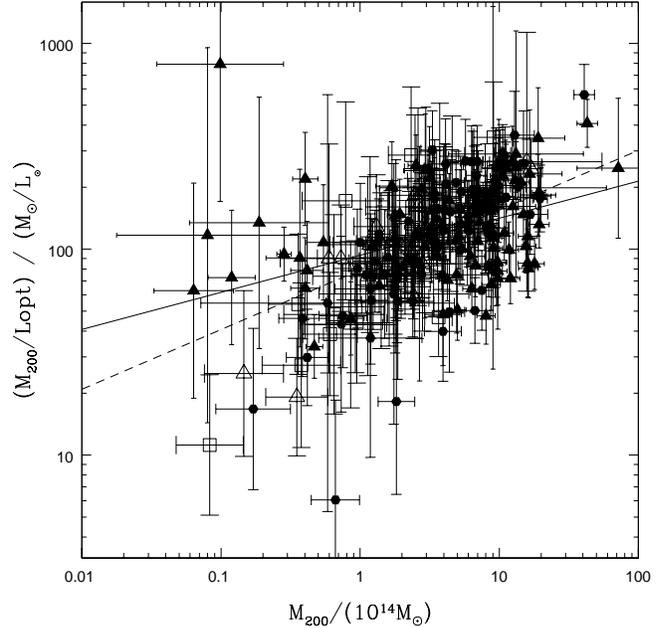}}
\end{minipage}
\end{center}
\caption{
The mass to light ratio versus the mass in the Sloan i band. The
filled points are the normal Abell clusters, the empty triangles are
the AXU clusters with a marginally significant X-ray detection, the
empty squares are the AXU clusters without X-ray detection, the filled
triangles are the X-ray selected clusters of the RASS-SDSS galaxy
cluster sample of paper III. The solid line is the best fit obtained
from the RASS-SDSS clusters, and the dashed line is the best fit
obtained from the Abell plus the RASS-SDSS clusters.}
\label{suml}
\end{figure}

Panel $a)$ of Fig. \ref{lom} shows the $L_{op}-M_{200}$ relation for
the RASS-SDSS sample.  Table \ref{table1} lists the best fit
parameters obtained performing a linear regression in the logarithmic
space. Note that the slope of the relations and their scatter are not
significantly different in other SDSS bands compared to the $i$-band.
Panels $b)$ and $c)$ of Fig. \ref{lom} show the location of the normal
Abell clusters and, respectively, of the AXU clusters, relatively to
the RASS-SDSS sample best fit line. Clearly, both the normal Abell
clusters and the AXU clusters obey the same $L_{op}-M_{200}$ relation
as the X-ray selected clusters.  The mean residual from the RASS-SDSS
relation is $\sim 0$ for both Abell clusters samples. Panel $d)$ of
Fig. \ref{lom} shows that adding the Abell clusters to the sample of
RASS-SDSS clusters does not alter the slope and the scatter of the
relation (see also Table \ref{table1}).  The slope of the
$L_{op}-M_{200}$ relation is confirmed to be smaller than
1. Therefore, we confirm the result of Paper III that the cluster
mass-to-light ratio $M/L$ is an increasing function of the cluster
mass, as shown in Fig. \ref{suml}.

\section{Nature of the AXU clusters}
As shown in the previous section, the AXU clusters are not a source of
scatter in the $L_{op}-M_{200}$ relation and, therefore, their optical
luminosity does not differ from that of the normal X-ray emitting
clusters of the same mass. On the other hand, they are significantly
offset from the $L_X-L_{op}$ and $L_X-M_{200}$ relations.  In this
section we try to elucidate the physical reason of this, in what
respect are the AXU clusters different from normal X-ray emitting
galaxy clusters. For this purpose, hereafter we compare the galaxy
luminosity functions, the relative fractions of red and blue galaxies,
galaxy number density profiles, and velocity distributions, of AXU and
normal clusters. We also look for the presence of optical
substructures, in order to see whether AXU clusters are more unrelaxed
systems than normal clusters.

\subsection{Luminosity functions}
\begin{figure*}
\begin{center}
\begin{minipage}{0.3\textwidth}
\resizebox{\hsize}{!}{\includegraphics{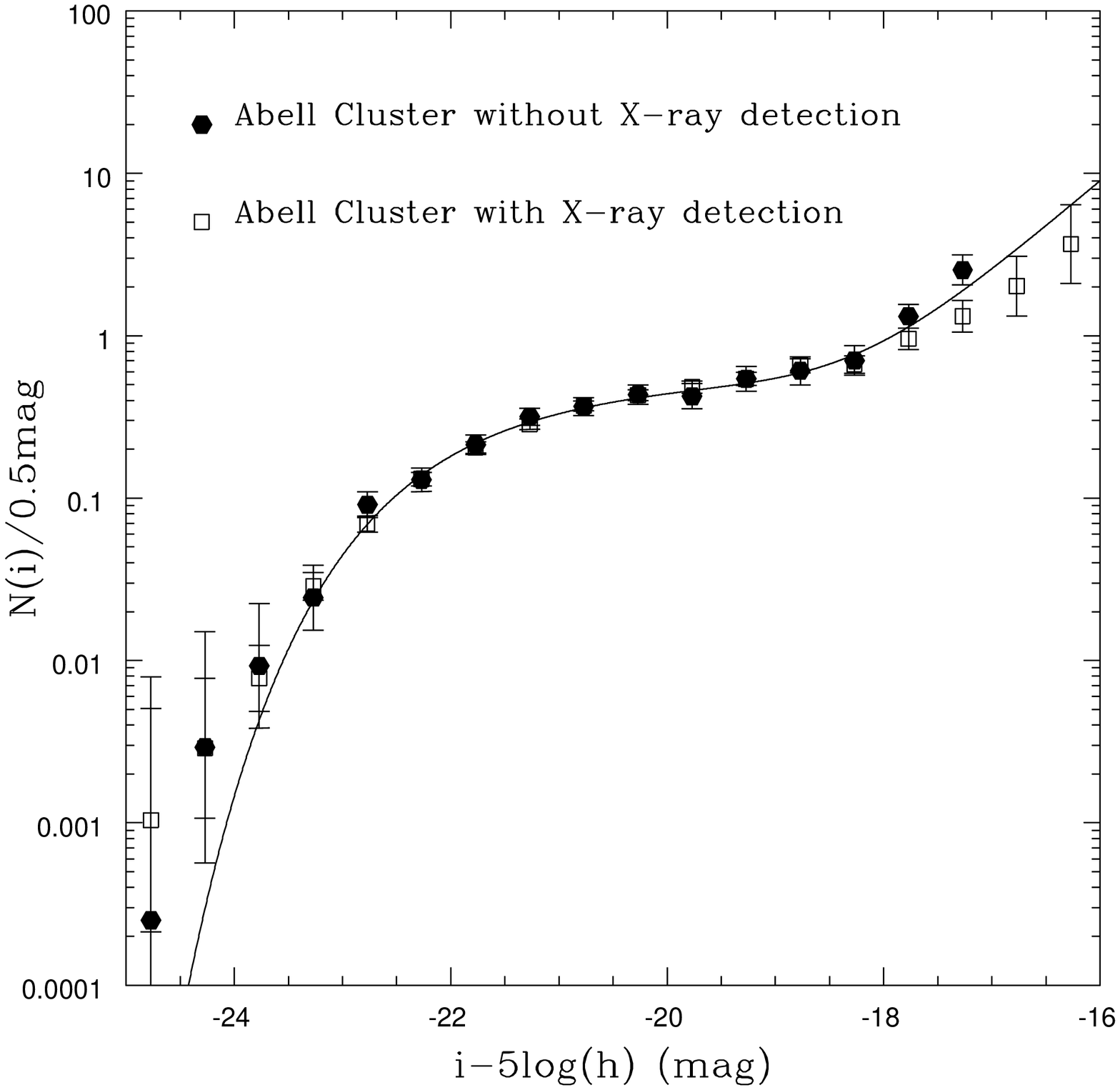}}
\end{minipage}
\begin{minipage}{0.3\textwidth}
\resizebox{\hsize}{!}{\includegraphics{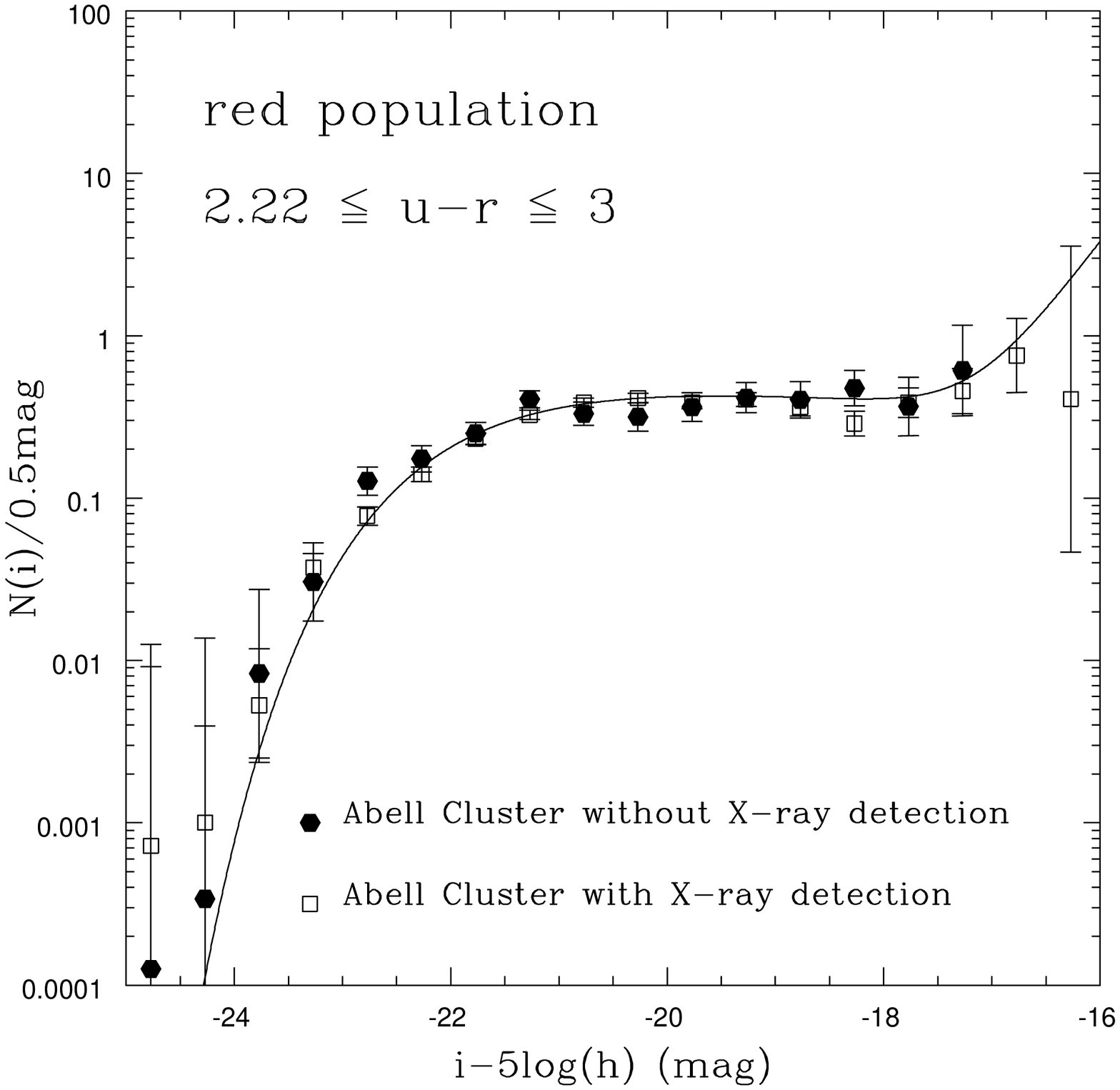}}
\end{minipage}
\begin{minipage}{0.3\textwidth}
\resizebox{\hsize}{!}{\includegraphics{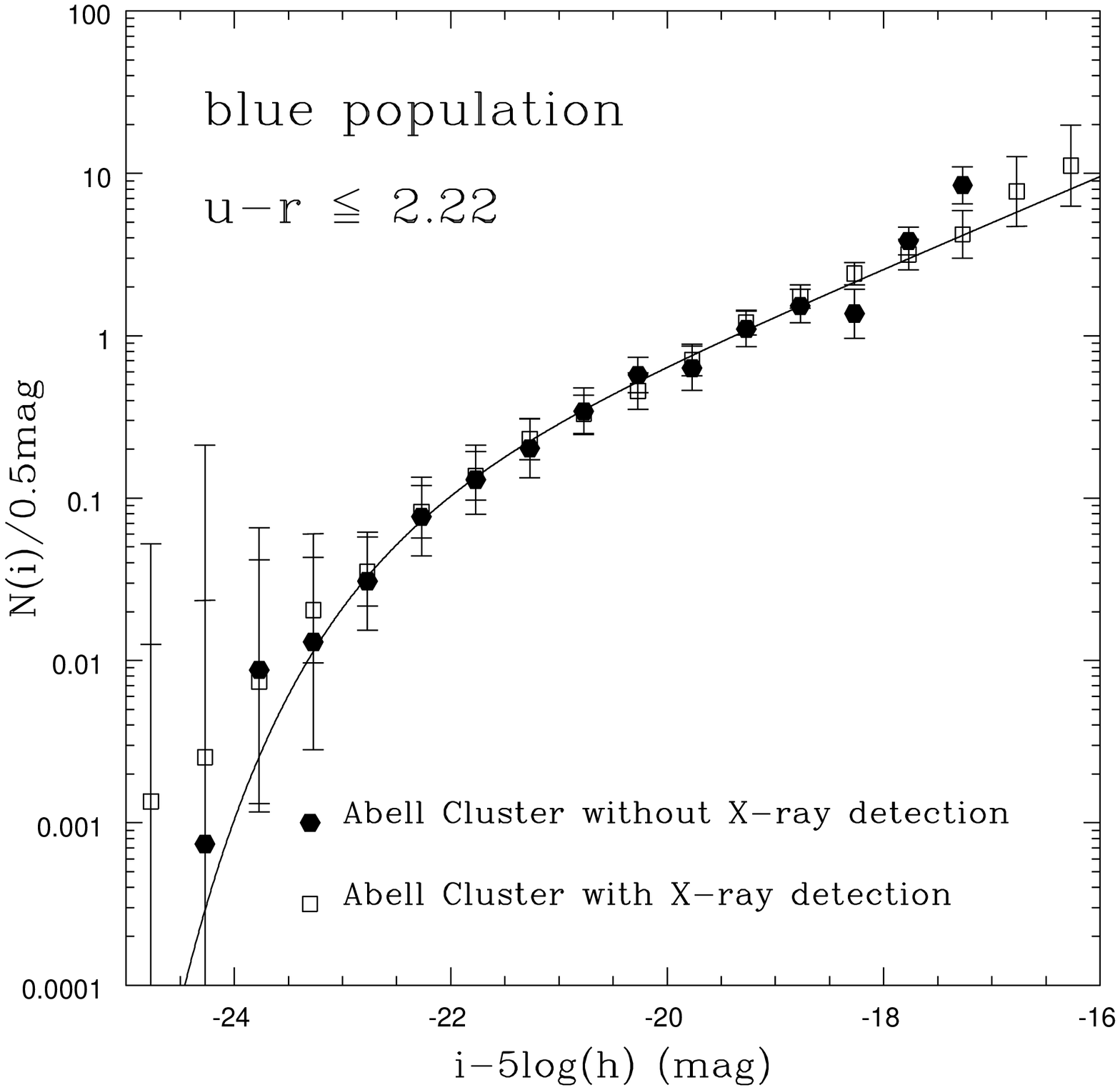}}
\end{minipage}
\end{center}
\caption{The luminosity function of the normal Abell clusters and the
AXU clusters. Left panel: composite cluster LFs of the whole galaxy
population; filled dots, AXU clusters; open squares, normal Abell
clusters; solid line, best-fit double Schechter LF obtained on the
X-ray selected RASS-SDSS cluster sample (see Paper IV). Middle panel:
same as left panel, but for the red galaxies only ($u-r \geq
2.22$). Right panel: same as left panel, but for the blue galaxies
only ($u-r<2.22$).}
\label{lf_xf}
\end{figure*}

We use the SDSS photometric data to compute a composite galaxy
luminosity function (LF) for the AXU systems, by stacking the
individual cluster LFs calculated within $r_{200}$. The individual LFs
are obtained by subtracting the field number counts calculated within
an annulus around the cluster, from the number counts in the cluster
region, as described in Paper II. In analogy to Paper IV, we have
distinguished between early and late type galaxies using a colour cut
at $u-r = 2.22$, as suggested by Strateva et al. (2001).
Fig. \ref{lf_xf} shows the composite LF of the AXU clusters (the
filled points) for the whole (left-hand panel), the red (middle
panel), and the blue (right-hand panel) cluster galaxy
populations. For comparison we also plot the corresponding composite
LFs of the normal Abell clusters (the empty squares), suitably
renormalised in order to ease the comparison with the LFs of the AXU
custers. The solid lines in the three panels of Fig. \ref{lf_xf} are
the best fit double Schechter functions, obtained in Paper IV on the
corresponding populations of the X-ray selected RASS-SDSS galaxy
clusters. It is clear that there are no significant differences among
the LFs of the three cluster samples, for any of the galaxy
populations. 

\subsection{Blue galaxy fractions}
\begin{figure}
\begin{center}
\begin{minipage}{0.5\textwidth}
\resizebox{\hsize}{!}{\includegraphics{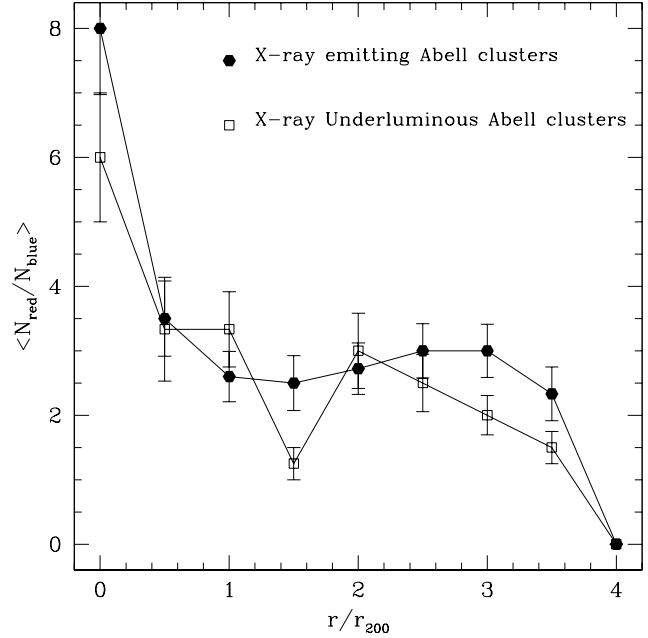}}
\end{minipage}
\end{center}
\caption{
The ratio of the numbers of red and blue cluster galaxies as a function of the
clustercentric distance in units of $r_{200}$.}
\label{red_sequence}
\end{figure}

In order to study the relative fraction of blue and red galaxies in
the different cluster samples, we stack together the galaxy colour
distributions of all the clusters of each given sample. Note that in
this case we only consider spectroscopically confirmed cluster
members, down to an absolute Petrosian magnitude $r_{Petro} \leq -20$,
and within $1.5 \, r_{200}$. We find that there is no difference
between the global colour distributions of the normal Abell clusters
and the AXU clusters.  AXU clusters do seem to have a larger fraction
of blue galaxies than normal Abell clusters in the outer regions (see
Fig. \ref{red_sequence}), but the statistical significance of this
difference is marginal.

\subsection{Galaxy number density profiles}
\begin{figure}
\begin{center}
\begin{minipage}{0.24\textwidth}
\resizebox{\hsize}{!}{\includegraphics{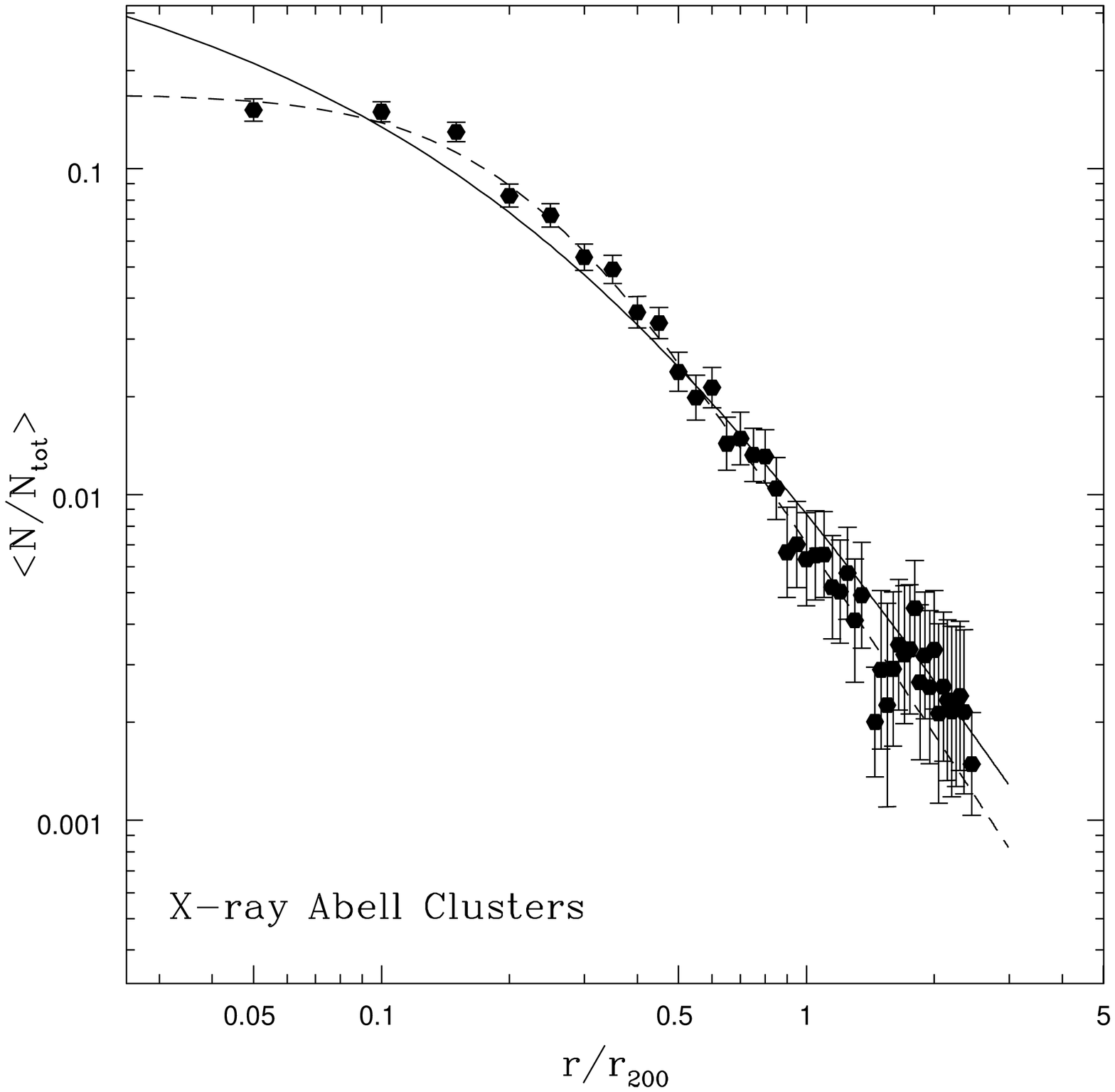}}
\end{minipage}
\begin{minipage}{0.24\textwidth}
\resizebox{\hsize}{!}{\includegraphics{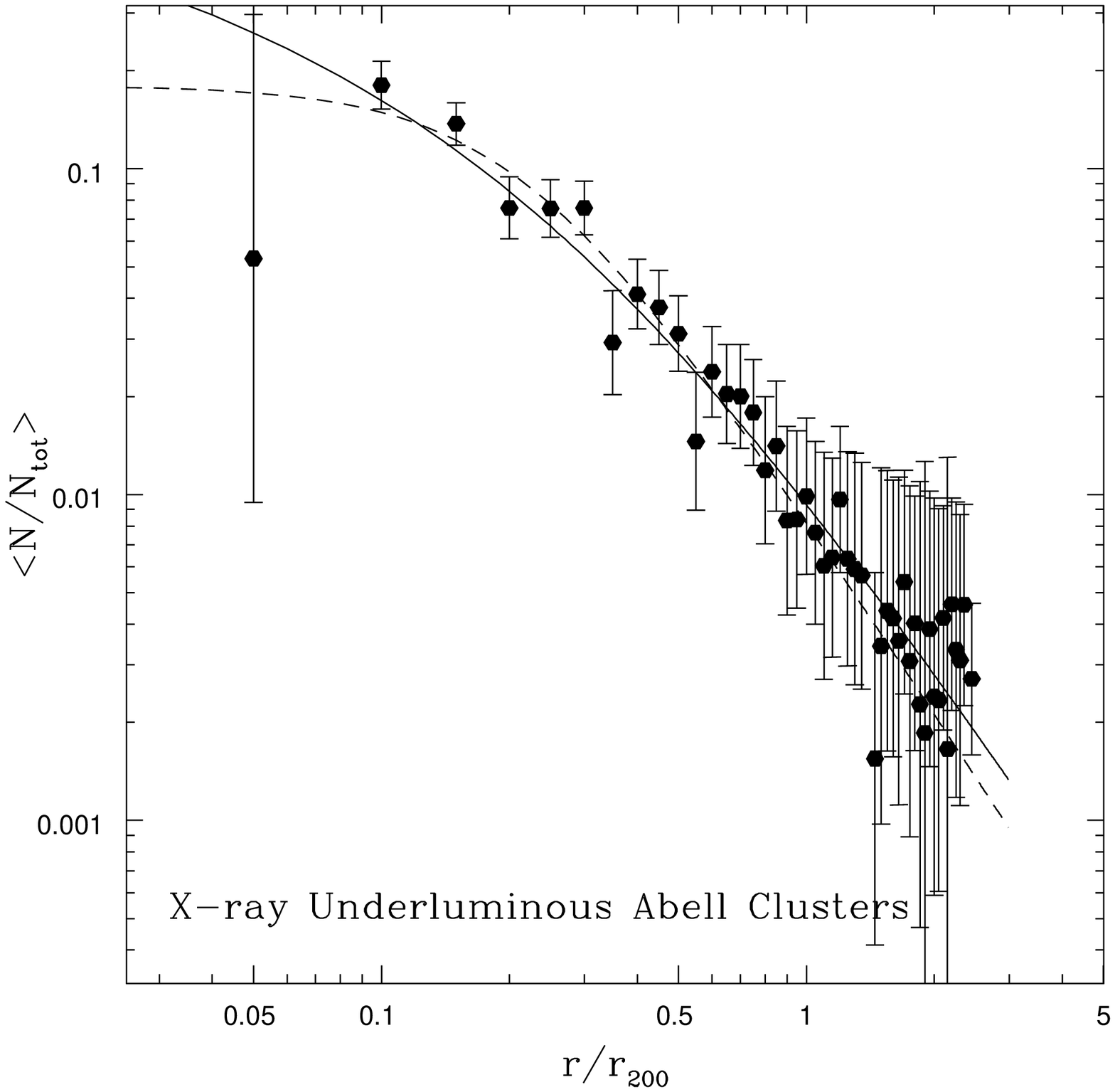}}
\end{minipage}
\end{center}
\caption{
The composite galaxy number density profiles of the normal
Abell clusters (left panel) and of the AXU clusters
(right panel).  The solid and dashed lines are the best fits given by a
projected NFW, and, respectively, a King density profile.}
\label{pro}
\end{figure}

In analogy to the analyses presented above, we compute the composite
galaxy number density profiles of the AXU clusters and the normal
Abell clusters. These are shown in Fig. \ref{pro}.  In order to
characterize these profiles, we fit two models to them. One is a King
(1966) profile, $\Sigma(x)=\Sigma_0 (1+x^2)^{-1}$, where $x=r/r_c$ and
$r_c$ is the core radius. The other model is the projected NFW
profile, which in 3-dimensions reads $n(x)=n_0 x^{-1}(1+x^2)^{-1}$,
where $x=c_g r/r_{200}$ and $c_g$ is the concentration parameter. The
surface density is then an integral of the three-dimensional profile
(see Bartelmann 1996 for more details).

For both the AXU and the normal Abell cluster samples the composite
radial profiles are better fit by a King profile (according to a
standard $\chi^2$ test). This is in agreement with previous results in
the literature (Adami et al. 1998b; D\'{\i}az et
al. 2005).  The best fit values of the core radii for the two samples
of clusters are $r_c/r_{200} = 0.209 \pm 0.006 \, \rm{Mpc}$ (normal Abell clusters)
and $r_c/r_{200} = 0.218 \pm 0.009 \, \rm{Mpc}$ (AXU clusters).  Therefore the two
profiles are perfectly consistent. However, we note that in the case
of the AXU clusters also a NFW profile provides an acceptable fit to
the data. This is however not due to a more cuspy profile than that of
the normal Abell clusters, but to the large error bar in the first bin
of the number density profile. Such a large error bar is due to a
paucity of galaxies in the very centre of AXU clusters. Hence AXU
clusters, relative to normal Abell clusters, seem to have a lower
central density of galaxies. This is consistent with their larger
fraction of blue galaxies (see the previous section) when we convolve
this information with the morphology-density relation (Dressler 1980).

\subsection{Galaxy velocity distributions}
In this subsection we analyze the composite galaxy velocity
distributions of the AXU clusters and the normal X-ray emitting
clusters. The differences between the mean cluster velocity and the
velocities of its member galaxies are normalized by $\sigma_c$, the
global cluster velocity dispersion. Each individual cluster velocity
distribution is then normalized to the total number of cluster members
in the considered cluster region.  We consider only member galaxies
with absolute Petrosian magnitude $r_{Petro} \leq -20$ mag, which is
brighter than any cluster limiting magnitude.  We estimate the
incompleteness of clusters spectroscopic samples by comparing the
number of cluster spectroscopic members found within $3.5 \,
r_{200}$ and within the chosen absolute magnitude limit, with the
number of cluster members obtained from the photometric data. The
photometric sample is not affected by incompleteness down to the
chosen magnitude limit. The number of photometric cluster members is
obtained by subtracting the number of background galaxies at the same
magnitude limit, rescaled by the cluster area, from the number of
galaxies (cluster$+$field) in the cluster region. From this analysis
we conclude that all the clusters have a spectroscopic completeness
$\geq 80$\% down to $r_{Petro} \leq -20$ mag.

\begin{figure*}
\begin{center}
\resizebox{\hsize}{!}{\includegraphics{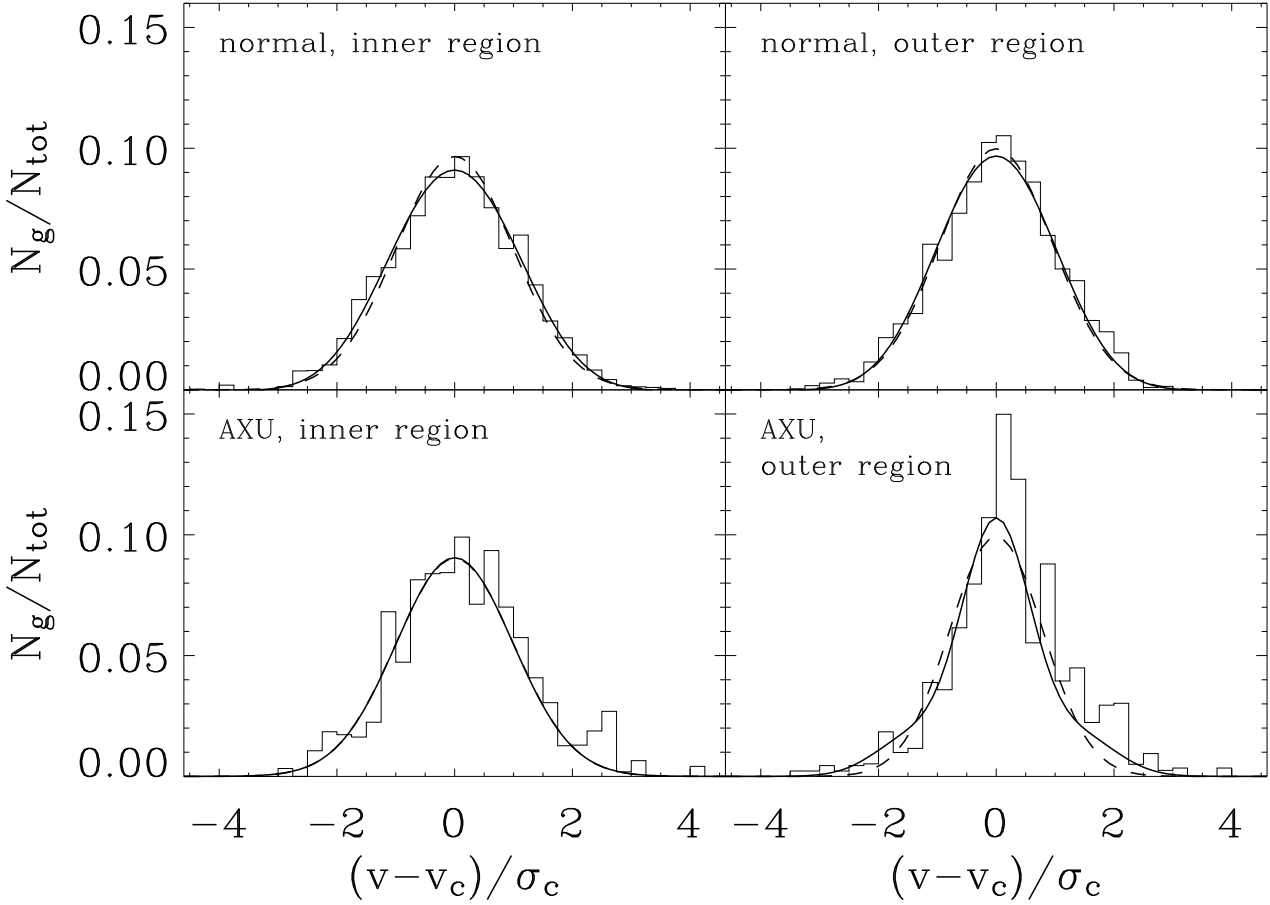}}
\end{center}
\caption{ The composite cluster velocity distributions. Top-left
panel: the velocity distribution of galaxies in normal Abell clusters
(histogram), within $1.5 \, r_{200}$. Top-right panel: the velocity
distribution of galaxies in normal Abell clusters (histogram), at
clustercentric distances in the range 1.5--$3.5 \,
r_{200}$. Bottom-left panel: the velocity distribution of galaxies in
AXU clusters (histogram), within $1.5 \, r_{200}$. Bottom-right panel:
the velocity distribution of galaxies in AXU clusters (histogram), at
clustercentric distances in the range 1.5--$3.5 \, r_{200}$. In each
panel, the dashed line represents the best-fit Gaussian, and the solid
line the best-fit obtained with a GH polynomial of order 4.}
\label{vdist-ghfit}
\end{figure*}

Fig. \ref{vdist-ghfit} shows the composite cluster velocity
distributions of the normal Abell clusters and the AXU clusters, for
two clustercentric distance intervals, $r/r_{200} \leq 1.5$ ('inner'
sample hereafter), and $1.5 < r/r_{200} \leq 3.5$ ('outer' sample
hereafter). The best-fit Gaussians are overplotted as dashed
lines. The best-fit Gaussian dispersion decreases from $1.00 \pm 0.01$
to $0.96 \pm 0.02$ from the inner to the outer velocity distribution
of the normal Abell clusters. The decrease is much stronger for the
AXU clusters, from $1.00 \pm 0.05$ to $0.80 \pm 0.07$. Hence, the
velocity dispersion profile is much steeper for the AXU clusters than
for the normal Abell clusters. It is reminiscent of the steep
velocity dispersion profile of late-type cluster galaxies (Biviano
et al. 1997; Adami et al. 1998c; Biviano \& Katgert 2004).  

In order to gain more insight into the meaning of this result, we
consider statistics that address the {\em shape} of the velocity
distributions. A classical shape estimator, the kurtosis, is not
recommended because it is very much influenced by the tails of the
distribution. Instead, we consider the more robust T.I.. The values of
the scaled tail index for the considered distribution are 1.05, 0.88,
1.16, 1.45 for the four subsamples (inner normal, outer normal, inner
AXU, outer AXU, respectively). As explained above, values larger than
unity indicate a leptokurtic distribution (i.e. more centrally peaked
than a Gaussian), while values smaller than unity indicate a
platikurtic distribution (i.e. more flat-topped than a Gaussian). Only
the scaled tail index value 1.45 is significantly different from unity
at $> 99$\% confidence level. We conclude that the outer velocity
distribution of the AXU clusters is not only significantly narrower
than all other velocity distributions, but it is also significantly
non-Gaussian, leptokurtic in particular.  Leptokurtic velocity
distributions occur in the outer cluster regions when the external
cluster members are characterized by radially elongated orbits
(Merritt 1987; van der Marel et al. 2000). Cosmological simulations
predict that halos should display leptokurtic velocity distributions
in their infall regions, characterized by ordered flows (Wojtak et
al. 2005).

In order to estimate the amount of radial anisotropy required to fit
the shape of the outer velocity distribution of AXU clusters, we
determine the value of the Gauss-Hermite (GH hereafter) moment of
order four (see, e.g., van der Marel et al. 2000). For completeness we
determine the GH moments also for the velocity distributions of the
other three subsamples. As expected from the T.I. analysis
above, the GH polynomial fits to the velocity distributions of the
normal Abell cluster galaxies, and of the inner AXU cluster galaxies,
are very similar to the Gaussian fits, and only for the velocity
distribution of the outer AXU cluster galaxies there is a clear
difference between the GH polynomial fit and the Gaussian fit (see
Fig. \ref{vdist-ghfit}).

We then compare the values of the $4^{th}$ GH moments of these
velocity distributions with the predictions of the dynamical models of
van der Marel et al. (2000, see their Figure 8). While these
predictions do depend on the number density distribution of the
considered galaxy population, such a dependence is not strong. Hence,
direct comparisons with van der Marel et al. dynamical models should
provide useful informations on the orbital anisotropy of the galaxy
populations.

The $4^{th}$ order GH moments are $-0.018$ and $-0.012$, for the inner
and outer velocity distributions of normal Abell cluster galaxies,
respectively, and $0.002$ and $0.106$ for the inner and outer velocity
distributions of AXU cluster galaxies, respectively. These values are
all consistent with isotropic orbits, except that of the outer
velocity distribution of the AXU cluster galaxies.  For this
population, we find $\sigma_r/\sigma_t \sim 2$ where $\sigma_r$ and
$\sigma_t$ are the radial and tangential velocity dispersions of the
galaxy population.

The analysis of the galaxy velocity distributions reveals a clear
difference between normal Abell clusters and AXU clusters. The
characteristics of the velocity distribution of AXU cluster galaxies
is reminiscent of an infalling galaxy population, such as the one seen
in numerical simulations in the external regions of dark matter haloes
(Wojtak et al. 2005). The higher fraction of blue galaxies seen in AXU
clusters, relative to that seen in normal Abell clusters, is certainly
consistent with a higher fraction of infalling galaxies, since these
must be part of the field galaxy population.

\begin{figure}
\begin{center}
\begin{minipage}{0.49\textwidth}
\resizebox{\hsize}{!}{\includegraphics{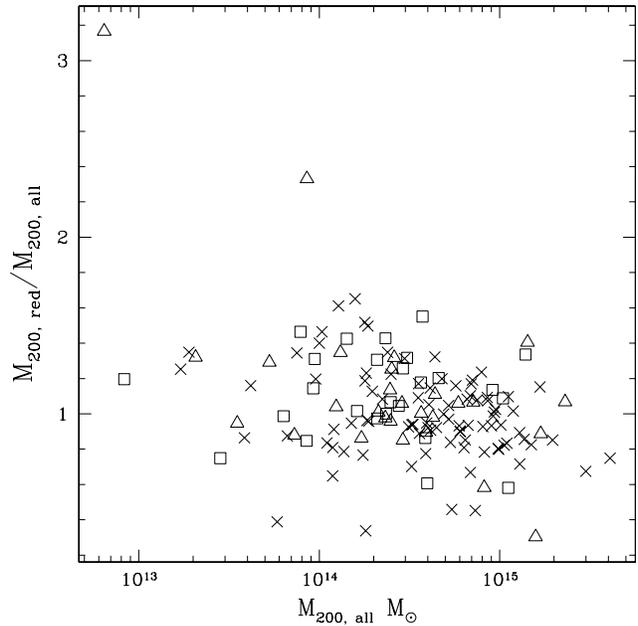}}
\end{minipage}
\end{center}
\caption{
Ratio of $M_{200}$ calculated only with the cluster red galaxy members
($M_{200,red}$) and $M_{200}$ calculated with all the cluster members
($M_{200,all}$) versus $M_{200,all}$. The crosses are the 'normal'
Abell clusters, the empty triangles and squares are the AXU systems
with marginally significant and without X-ray detection,
respectively. }
\label{lowlx}
\end{figure}

\section{Discussion and Conclusions}
We have studied the X-ray and optical properties of 137 isolated Abell
clusters. Each object has a confirmed three-dimensional overdensity of
galaxies. We have looked for the X-ray counterpart of each system in
the RASS data. Three classes of objects have been identified, where
the classification is based on the quality of the X-ray detection. 86
clusters out of the 137 Abell systems have a clear X-ray detection and
are considered normal X-ray emitting clusters (the 'normal Abell
clusters'). 27 systems have a X-ray detection of low significance
(less the $3\sigma$) and 24 do not have clear X-ray detection (a rough
estimate of $L_X$ is provided but with huge statistical errors). 

The normal Abell clusters follow the same scaling relations observed
in the X-ray selected RASS-SDSS clusters. The $24+27$ Abell clusters
with unsecure X-ray detection appear to be outliers in the
$L_X-M_{200}$ relation determined for X-ray luminous clusters. Their
X-ray luminosity is on average one order of magnitude fainter than
would be expected for their mass . A careful analysis of the 3D
galaxy overdensity of these systems reveals that the individual galaxy
velocity distributions in the virial region are gaussian in 90\% of
the clusters and are not ascribable to the superposition of smaller
interacting systems. We conclude that these Abell cluster with
unsecure X-ray detection in RASS are not spurious detections in the
redshift distribution, but are a distinct class of objects. Due to
their location with regard to the RASS-SDSS $M-L_X$ relation we call
them 'Abell X-ray underluminous clusters' or AXU clusters for
short. Several AXU clusters are confirmed to be very faint X-ray
objects in the literature. Their X-ray flux is probably too low to be
detected in the RASS survey.  Yet, AXU clusters are not outliers from
the $L_{op}-M_{200}$ relation, i.e. they have a normal optical
luminosity given their mass. Hence, the distinctive signature of AXU
clusters seems to lie in an X-ray luminosity which is unexpectedly
low.

We have looked for other properties of AXU clusters that make them
different from normal Abell clusters. We have shown that AXU clusters
do not have more substructures than normal Abell clusters. The galaxy
luminosity functions within the virial region of the two cluster
samples are very similar to each other. Rather similar are their
galaxy number density profiles, even if the AXU clusters seem to lack
galaxies near the core, relative to normal Abell clusters (but the
significance of this result is low). The fractions of blue galaxies in
the two kinds of clusters are only marginaly different, AXU clusters
being characterized by a higher fraction.

The main difference between the two classes of objects lies in the
velocity distribution of their member galaxies. The galaxy velocity
distribution of the normal Abell clusters is perfectly fitted by a
Gaussian both in the inner, virialized region ($\le 1.5 \, r_{200}$),
and also in the external region ($1.5 \, r_{200} \le r \le 3.5 \,
r_{200}$). The AXU clusters instead have a Gaussian velocity
distribution only within the virial region. In the external region,
their velocity distribution is significantly more peaked than a
Gaussian. The analysis of its shape by comparison with dynamical
models available in the literature (van der Marel et al. 2000),
suggests a radially anisotropic galaxy orbital distribution. However,
the galaxies in this external region need not be in dynamical
equilibrium with the cluster potential. As a matter of fact, a
leptokurtic shape of the velocity distribution is a typical signature
of the external, infall regions of dark matter haloes (Wojtak et al.
2005).

The analysis of the velocity distribution of the AXU clusters in their
outer regions hence suggests the presence of an unvirialized component
of the galaxy population, still in the process of accretion onto the
cluster. This infalling population would be mainly composed of field,
hence blue, galaxies, which could then explain the excess of blue
galaxies in AXU clusters, relative to normal Abell clusters.  On the
other hand, the Gaussian velocity distribution in the inner region
suggests that there the galaxy population is dynamically more evolved,
and probably virialized.

By a similar analysis on a different sample of X-ray underluminous
clusters, Bower et al. (1997) came to propose two different
scenarios. AXU clusters could be severely affected by projection
effects arising from surrounding large-scale structure filaments
elongated along the line-of-sight.  Their velocity dispersion, and
hence their virial masses would then be severely overestimated by
interlopers in the filaments. In the alternative scenario AXU clusters
could be clusters not yet formed, or in the phase of forming, or, at
least, caught at a particular stage of their evolution, while they are
undergoing a rapid mass growth.

Should the former of the two scenarios apply, we would expect AXU
clusters to be X-ray underluminous for their mass, but they could
still be optically luminous because we partly see the light of the
filament projected onto the cluster. However, contamination by
interlopers does affect the optical luminosity estimate, but not so
much as the virial mass estimate, and not so much in the $i$ band,
where contamination by the field (hence blue) galaxies should be
small. Therefore, in such scenario it would be surprising that the
clusters obey so well the $L_{op}-M_{200}$ relation, which requires
that the effects of the filament on the dynamical mass estimate and
the optical light in the aperture both conspire not to produce an
offset from the relation. It would also be surprising that the AXU
clusters show a galaxy LF perfectely consistent with the steep LF
found in galaxy clusters (see Popesso et al. paper II and IV) and not
the flat LF observed in the field (Blanton et al. 2005). Instead AXU
clusters are not outliers from the $L_{op}-M_{200}$ relation. If
anything, AXU clusters are overluminous in the optical for their
mass. In fact, the biweight-average (see Beers et al. 1990) $i$-band
mass-to-light ratios of normal Abell clusters and AXU clusters are
$150 \pm 10 \, M_{\odot}/L_{\odot}$, and $110 \pm 10 \,
M_{\odot}/L_{\odot}$, respectively.

As a further test, we have re-calculated the virial masses of all
clusters by considering only red cluster members belonging to the red
sequence in the $u-i$ vs. $i$ color-magnitude diagram. In this way
contamination by interlopers is strongly reduced (see, e.g., Biviano
et al. 1997).  Masses computed using all cluster members are compared
to masses computed using only red-sequence members in
Fig. \ref{lowlx}. The cluster masses do not change significantly when
only red-sequence members are used to calculate them, suggesting a low
level of contamination by interlopers.

The results of our analyses therefore support Bower et al.'s
alternative scenario, namely AXU clusters are systems in the stage of
formation and/or of significant mass accretion.  If AXU clusters are
still forming, the intra-cluster gas itself may still be infalling or
have not yet reached the virial temperature. In addition, for AXU
clusters undergoing massive accretion, it is to some degree possible
that the continuous collisions of infalling groups is affecting the
gas distribution, lowering its central density (such as in the case of
the so called 'bullet cluster', see Barrena et al. 2002 and Clowe et
al. 2004). In both cases the X-ray luminosity would be substantially
lower than predicted for the virial mass of the system, because of its
dependence on the square of the gas density. We note however that a
virialized cluster undergoing a strong collision with an infalling
group would show up as a substructured cluster, yet the AXU clusters
do not show an increased level of substructures when compared to
normal Abell clusters. In summary, we know that the X-ray emission is
very much dominated by the central region whereas the optical
properties are more global.  Therefore it could well be that we see a
rough relaxation on the large scale (within $1.5r_{200}$) of the
galaxy system reflected by a rough Gaussian galaxy velocity
distribution, while the central region has not yet settled to reach
the high density and temperatures of the luminous X-ray clusters.

In order to explore this further, we need much more detailed
information on the distribution of the density and temperature of the
intracluster gas in AXU clusters, something that cannot be done with
the RASS data, but requires the spatial resolution and sensitivity of
XMM-Newton.

Our results give supports to the conclusion of Donahue et al. (2002)
concerning the biases inherent in the selection of galaxy clusters in
different wavebands. While the optical selection is prone to
substantial projection effects, also the X-ray selection is not
perfect or not simple to characterize. The existence of X-ray
underluminous clusters, even with large masses, makes it difficult to
reach the needed completeness in mass for cosmological
studies. Moreover, as discussed in Paper III, the relation between the
X-ray luminosity and mass is not very tight even for the X-ray bright
clusters, and the relation between cluster masses and optical
luminosities is as tight or perhaps even tighter. Clearly, a
multi-waveband approach is needed for optimizing the completeness and
reliability of clusters samples.

On the other hand, it becomes clear that for precision cosmology we
also need a more observationally oriented prescription of cluster
selection from theory, rather than a mere counting of "relaxed" dark
matter halos.  Predicted distribution functions closer to the
observational parameters like temperature or velocity dispersion
distribution functions and their relations to X-ray and optical
luminosity are needed.

\vspace{2cm}
We thank the referee F. Castander for the useful comments, which
helped in improving the paper. We thank Alain Mazure for useful
discussion.  Funding for the creation and distribution of the SDSS
Archive has been provided by the Alfred P.  Sloan Foundation, the
Participating Institutions, the National Aeronautics and Space
Administration, the National Science Foundation, the U.S.  Department
of Energy, the Japanese Monbukagakusho, and the Max Planck
Society. The SDSS Web site is http://www.sdss.org/. The SDSS is
managed by the Astrophysical Research Consortium (ARC) for the
Participating Institutions.  The Participating Institutions are The
University of Chicago, Fermilab, the Institute for Advanced Study, the
Japan Participation Group, The Johns Hopkins University, Los Alamos
National Laboratory, the Max-Planck-Institute for Astronomy (MPIA),
the Max-Planck-Institute for Astrophysics (MPA), New Mexico State
University, University of Pittsburgh, Princeton University, the United
States Naval Observatory, and the University of Washington.

\clearpage
\onecolumn

\begin{appendix}
\section{The Abell Cluster Catalog}
Here we list the properties of the 137 spectroscopically confirmed
Abell systems extracted from the SDSS DR3, used in this paper.
The meaning of the individual columns is the following:
\begin{itemize}  
\item [-] column 1: the name of the Abell cluster  
\item [-] column 2: the number of cluster members within 1 Abell radius
\item [-] column 3: the cluster mean redshift
\item [-] column 4: the cluster velocity dispersion and its error
\item [-] column 5: the cluster mass within $r_{200}$, $M_{200}$,
in units of $10^{14} M_{\odot}$
\item [-] column 6:  the cluster mass within $r_{200}$ calculated using only the cluster red members, $M_{200, \rm{red}}$, in units of $10^{14} M_{\odot}$; $M_{200, \rm{red}}$ is given only for the clusters with at least 10 red members
\item [-] column 7: the fractional error on $M_{200}$ and $M_{200, \rm{red}}$
\item [-] column 8: the cluster virial radius, $r_{200}$, in Mpc
\item [-] column 9: the cluster optical luminosity $L_{op}$ and its error, 
in unit of $10^{12}L_{\odot}$
\item [-] column 10: the cluster X-ray luminosity in the ROSAT energy band 
(0.1-2.4 erg $\rm{s}^{-1}$), in unit of $10^{44}$ erg $\rm{s}^{-1}$
\item [-] column 11: the fractional error on the X-ray luminosity
\item [-] column 12: the Dressler \& Shectman probability that a cluster
does not contain substructures, $P_DS$
(values $<0.1$ indicate clusters that are likely to contain substructures)
\item [-] column 13: the X-ray class: 0 for the normal X-ray emitting cluster, 1 for the Abell systems with less the $3\sigma$ X-ray detection, 2 for the X-ray non-detected Abell Clusters.
\end{itemize}

\begin{longtable}{ccccccccccccc}\hline\hline
Name & $N_{mem}$ & $z_c$ & $\sigma_c$ & $M_{200}$ &  $M_{200, \rm{red}}$ & $er_{M}$ & $r_{200}$ & $L_{op}$ &
$L_X$ & $er_{L_X}$ & $P_{DS}$ & X-class \\
\hline
\endhead
 a0116 &   24 & 0.0661&  582 $\pm$ 73  &  4.66 &     4.71  &  0.29 &  1.6 &  1.59 $\pm$  0.35 &  0.090 &  0.25 & 0.19 & 0 \\
 a0117 &   95 & 0.0550&  559 $\pm$ 41  &  4.03 &     3.85  &  0.15 &  1.5 &  2.52 $\pm$  0.64 &  0.064 &  0.26 & 0.96 & 0 \\
 a0129 &   19 & 0.1501&  749 $\pm$ 119 &  8.78 &     9.80  &  0.36 &  2.0 &  5.03 $\pm$  1.87 &  0.548 &  0.26 & 0.62 & 0 \\
 a0130 &   21 & 0.1106&  447 $\pm$ 84  &  2.81 &     2.34  &  0.37 &  1.4 &  2.12 $\pm$  0.48 &  0.161 &  0.61 & 0.13 & 1 \\
 a0152 &   69 & 0.0589&  729 $\pm$ 59  &  7.04 &     5.12  &  0.19 &  1.8 &  2.38 $\pm$  0.52 &  0.057 &  0.25 & 0.09 & 0 \\
 a0168 &  110 & 0.0450&  559 $\pm$ 36  &  3.57 &     2.89  &  0.14 &  1.5 &  2.70 $\pm$  0.36 &  0.370 &  0.09 & 0.57 & 0 \\
 a0175 &   37 & 0.1285&  606 $\pm$ 60  &  4.47 &     3.73  &  0.31 &  1.6 &  9.88 $\pm$  2.53 &  1.114 &  0.29 & 0.84 & 0 \\
 a0190 &   17 & 0.1021&  431 $\pm$ 122 &  2.19 &     0.61  &  0.66 &  1.2 &  1.64 $\pm$  0.41 &  0.034 &  0.70 & 0.00 & 0 \\
 a0208 &   31 & 0.0793&  499 $\pm$ 60  &  2.60 &     2.42  &  0.29 &  1.3 &  1.64 $\pm$  0.32 &  0.237 &  0.22 & 0.68 & 0 \\
 a0243 &   32 & 0.1125&  469 $\pm$ 50  &  2.52 &     2.04  &  0.30 &  1.3 &  2.83 $\pm$  0.39 &  0.000 &  0.00 & 0.40 & 1 \\
 a0315 &   16 & 0.1740&  636 $\pm$ 96  &  6.61 &     6.24  &  0.41 &  1.8 &  4.91 $\pm$  0.91 &  0.056 &  1.20 & 0.50 & 2 \\
 a0351 &   14 & 0.1108&  510 $\pm$ 118 &  2.70 &     2.27  &  0.42 &  1.3 &  2.35 $\pm$  0.57 &  0.016 &  1.50 & 0.99 & 2 \\
 a0412 &   31 & 0.1092&  585 $\pm$ 50  &  3.29 &     3.04  &  0.28 &  1.4 &  2.20 $\pm$  0.69 &  0.071 &  0.53 & 0.11 & 2 \\
 a0441 &   25 & 0.1443&  907 $\pm$ 554 & 17.17 &     4.80  &  1.25 &  2.5 &  5.95 $\pm$  1.24 &  0.218 &  0.44 & 0.00 & 2 \\
 a0607 &   34 & 0.0962&  501 $\pm$ 88  &  2.88 &     2.65  &  0.37 &  1.4 &  2.61 $\pm$  0.42 &  0.023 &  0.57 & 0.56 & 1 \\
 a0620 &   14 & 0.1323&  518 $\pm$ 76  &  2.17 &     2.78  &  0.61 &  1.2 &  3.34 $\pm$  0.91 &  0.787 &  0.17 & 0.35 & 0 \\
 a0626 &   15 & 0.1168&  757 $\pm$ 158 &  7.16 &     5.48  &  0.40 &  1.8 &  3.30 $\pm$  0.83 &  0.092 &  0.44 & 0.32 & 0 \\
 a0628 &   61 & 0.0834&  642 $\pm$ 63  &  5.98 &     4.46  &  0.19 &  1.7 &  2.81 $\pm$  0.44 &  0.208 &  0.35 & 0.81 & 0 \\
 a0631 &   48 & 0.0826&  577 $\pm$ 48  &  3.77 &     3.11  &  0.19 &  1.5 &  1.09 $\pm$  0.27 &  0.061 &  0.42 & 0.49 & 0 \\
 a0646 &   29 & 0.1266&  738 $\pm$ 96  & 10.45 &     9.72  &  0.24 &  2.1 &  3.82 $\pm$  0.80 &  2.487 &  0.09 & 0.92 & 0 \\
 a0655 &   47 & 0.1276&  736 $\pm$ 78  &  9.47 &     9.21  &  0.20 &  2.0 &  7.29 $\pm$  0.94 &  2.527 &  0.16 & 0.88 & 0 \\
 a0660 &   26 & 0.0642&  752 $\pm$ 138 &  7.91 &     7.61  &  0.43 &  1.9 &  1.62 $\pm$  0.32 &  0.000 &  0.00 & 0.27 & 2 \\
 a0667 &   17 & 0.1441&  512 $\pm$ 85  &  2.05 &     1.33  &  1.25 &  1.2 &  2.88 $\pm$  0.63 &  1.998 &  0.11 & 0.53 & 0 \\
 a0682 &   17 & 0.1147&  266 $\pm$ 242 &  0.75 &     0.22  &  2.12 &  0.9 &  1.07 $\pm$  0.25 &  0.057 &  0.50 & 0.57 & 0 \\
 a0685 &   16 & 0.1464&  496 $\pm$ 56  &  4.47 &     3.49  &  0.26 &  1.6 &  3.46 $\pm$  0.88 &  0.000 &  0.00 & 0.07 & 2 \\
 a0714 &   29 & 0.1392&  574 $\pm$ 78  &  4.97 &     4.86  &  0.26 &  1.6 &  4.16 $\pm$  1.10 &  0.041 &  0.83 & 0.86 & 2 \\
 a0716 &   17 & 0.1188&  494 $\pm$ 144 &  2.88 &     2.37  &  0.59 &  1.4 &  1.36 $\pm$  0.43 &  0.009 &  1.50 & 0.46 & 2 \\
 a0729 &   28 & 0.0978&  688 $\pm$ 87  &  3.38 &     3.90  &  0.36 &  1.4 &  1.21 $\pm$  0.40 &  0.232 &  0.22 & 0.19 & 0 \\
 a0733 &   11 & 0.1156&  392 $\pm$ 78  &  0.91 &      --   &  0.69 &  0.9 &  1.69 $\pm$  0.51 &  0.535 &  0.62 & 0.49 & 0 \\
 a0736 &   42 & 0.0619&  826 $\pm$ 98  & 10.11 &     9.08  &  0.29 &  2.1 &  4.66 $\pm$  0.77 &  0.061 &  0.25 & 0.04 & 0 \\
 a0847 &   16 & 0.1508&  704 $\pm$ 115 &  5.03 &     4.11  &  0.40 &  1.6 &  2.94 $\pm$  0.54 &  0.730 &  0.21 & 0.87 & 0 \\
 a0856 &   19 & 0.1393&  450 $\pm$ 69  &  1.92 &     2.61  &  0.48 &  1.2 &  1.74 $\pm$  0.45 &  0.407 &  0.38 & 0.01 & 0 \\
 a0860 &   31 & 0.0965&  941 $\pm$ 95  & 12.98 &     12.0  &  0.36 &  2.2 &  2.13 $\pm$  0.46 &  0.313 &  0.23 & 0.00 & 0 \\
 a0861 &   17 & 0.1259&  468 $\pm$ 104 &  3.29 &     2.88  &  0.44 &  1.4 &  2.51 $\pm$  0.51 &  0.237 &  0.27 & 0.67 & 1 \\
 a0866 &   10 & 0.1435&  266 $\pm$ 106 &  0.83 &      --   &  0.84 &  0.9 &  1.34 $\pm$  0.39 &  0.143 &  0.47 & 0.12 & 1 \\
 a0869 &   12 & 0.1198&  381 $\pm$ 127 &  1.74 &     2.02  &  0.66 &  1.2 &  1.81 $\pm$  0.39 &  0.241 &  0.30 & 0.29 & 1 \\
 a0892 &   23 & 0.0943&  470 $\pm$ 148 &  1.45 &     0.76  &  0.86 &  1.1 &  3.20 $\pm$  1.56 &  0.175 &  0.26 & 0.09 & 0 \\
 a0912 &   28 & 0.0906&  590 $\pm$ 82  &  3.72 &     3.05  &  0.31 &  1.5 &  3.01 $\pm$  0.60 &  0.021 &  0.62 & 0.75 & 0 \\
 a0917 &   11 & 0.1370&  403 $\pm$ 76  &  0.76 &      --   &  0.46 &  0.9 &  1.57 $\pm$  0.37 &  0.252 &  0.31 & 0.24 & 1 \\
 a0919 &   12 & 0.0954&  136 $\pm$ 37  &  0.21 &      --   &  0.66 &  0.6 &  0.59 $\pm$  0.16 &  0.033 &  0.55 & 0.18 & 2 \\
 a0933 &   56 & 0.0965&  455 $\pm$ 46  &  2.86 &     3.22  &  0.20 &  1.4 &  4.28 $\pm$  0.89 &  0.387 &  0.21 & 0.87 & 0 \\
 a0975 &   14 & 0.1192&  208 $\pm$ 58  &  0.48 &      --   &  0.50 &  0.8 &  0.67 $\pm$  0.19 &  0.068 &  0.50 & 0.96 & 1 \\
 a1038 &   13 & 0.1275&  253 $\pm$ 48  &  0.55 &     0.48  &  0.35 &  0.8 &  1.40 $\pm$  0.27 &  0.108 &  0.44 & 0.83 & 0 \\
 a1064 &   17 & 0.1318&  485 $\pm$ 93  &  2.30 &     2.21  &  0.34 &  1.3 &  2.39 $\pm$  0.59 &  0.211 &  0.33 & 0.74 & 0 \\
 a1066 &  100 & 0.0690&  731 $\pm$ 52  &  6.63 &     5.55  &  0.15 &  1.8 &  4.15 $\pm$  0.60 &  0.657 &  0.17 & 0.35 & 0 \\
 a1072 &   11 & 0.1173&  364 $\pm$ 83  &  1.45 &      --   &  0.67 &  1.1 &  1.12 $\pm$  0.36 &  0.029 &  0.86 & 0.10 & 2 \\
 a1076 &   18 & 0.1168&  418 $\pm$ 77  &  1.57 &     2.06  &  0.37 &  1.1 &  1.46 $\pm$  0.30 &  0.295 &  0.21 & 0.96 & 0 \\
 a1078 &   11 & 0.1242&  249 $\pm$ 51  &  0.50 &      --   &  0.66 &  0.8 &  1.38 $\pm$  0.38 &  0.182 &  0.37 & 0.60 & 1 \\
 a1092 &   26 & 0.1058&  449 $\pm$ 64  &  2.07 &     1.47  &  0.34 &  1.2 &  1.55 $\pm$  0.35 &  0.000 &  0.00 & 0.02 & 2 \\
 a1107 &   15 & 0.1508&  792 $\pm$ 104 & 10.03 &     10.3  &  0.33 &  2.1 &  2.61 $\pm$  0.78 &  0.265 &  0.38 & 0.03 & 1 \\
 a1132 &   27 & 0.1358&  880 $\pm$ 138 &  8.90 &     7.48  &  0.36 &  2.0 &  6.37 $\pm$  0.95 &  3.038 &  0.07 & 0.45 & 0 \\
 a1139 &   89 & 0.0395&  376 $\pm$ 34  &  1.68 &     1.07  &  0.19 &  1.2 &  1.16 $\pm$  0.25 &  0.136 &  0.16 & 0.24 & 0 \\
 a1143 &   13 & 0.1379&  459 $\pm$ 86  &  2.10 &      --   &  0.45 &  1.2 &  2.25 $\pm$  0.56 &  0.030 &  0.80 & 0.10 & 2 \\
 a1164 &   19 & 0.1057&  609 $\pm$ 144 &  4.24 &     5.79  &  0.54 &  1.6 &  1.75 $\pm$  0.60 &  0.074 &  0.61 & 0.15 & 1 \\
 a1171 &   16 & 0.0577&  161 $\pm$ 40  &  0.12 &     0.09  &  0.56 &  0.5 &  0.74 $\pm$  0.17 &  0.024 &  0.55 & 0.01 & 1 \\
 a1189 &   37 & 0.0969&  654 $\pm$ 196 &  4.58 &     4.26  &  0.59 &  1.6 &  2.63 $\pm$  0.44 &  0.087 &  0.32 & 0.20 & 0 \\
 a1205 &   80 & 0.0761&  865 $\pm$ 73  & 11.99 &     9.13  &  0.19 &  2.2 &  3.88 $\pm$  0.54 &  0.976 &  0.09 & 0.01 & 0 \\
 a1218 &   23 & 0.0801&  364 $\pm$ 75  &  0.91 &     0.63  &  0.39 &  0.9 &  0.81 $\pm$  0.17 &  0.017 &  0.56 & 0.58 & 2 \\
 a1221 &   11 & 0.1103&  289 $\pm$ 132 &  0.77 &      --   &  0.82 &  0.9 &  0.66 $\pm$  0.32 &  0.000 &  0.00 & 0.66 & 2 \\
 a1236 &   38 & 0.1021&  533 $\pm$ 59  &  3.72 &     2.27  &  0.29 &  1.5 &  2.57 $\pm$  0.45 &  0.150 &  0.30 & 0.06 & 0 \\
 a1302 &   47 & 0.1153&  691 $\pm$ 80  &  7.35 &     7.14  &  0.25 &  1.9 &  3.61 $\pm$  1.08 &  1.307 &  0.09 & 0.72 & 0 \\
 a1346 &   74 & 0.0983&  709 $\pm$ 54  &  7.64 &     4.59  &  0.18 &  1.9 &  3.98 $\pm$  0.59 &  0.208 &  0.30 & 0.15 & 0 \\
 a1364 &   41 & 0.1066&  553 $\pm$ 59  &  2.85 &     2.80  &  0.24 &  1.4 &  4.27 $\pm$  0.94 &  0.040 &  0.80 & 0.55 & 2 \\
 a1366 &   42 & 0.1164&  691 $\pm$ 70  &  7.72 &     8.13  &  0.20 &  1.9 &  2.61 $\pm$  0.53 &  1.550 &  0.10 & 0.88 & 0 \\
 a1368 &   27 & 0.1293&  735 $\pm$ 92  &  7.84 &     8.40  &  0.28 &  1.9 &  3.71 $\pm$  0.65 &  0.130 &  0.47 & 0.06 & 0 \\
 a1376 &   16 & 0.1179&  461 $\pm$ 204 &  3.11 &     3.43  &  0.88 &  1.4 &  1.62 $\pm$  0.39 &  0.013 &  0.67 & 0.31 & 2 \\
 a1387 &   35 & 0.1310&  692 $\pm$ 73  &  6.73 &     5.44  &  0.27 &  1.8 &  4.91 $\pm$  0.66 &  0.693 &  0.20 & 0.05 & 0 \\
 a1392 &   11 & 0.1361&  517 $\pm$ 146 &  3.86 &      --   &  0.61 &  1.5 &  3.11 $\pm$  0.71 &  0.707 &  0.19 & 0.49 & 0 \\
 a1399 &   23 & 0.0910&  251 $\pm$ 59  &  0.46 &     0.33  &  0.52 &  0.8 &  1.83 $\pm$  0.27 &  0.000 &  0.00 & 0.03 & 2 \\
 a1406 &   14 & 0.1170&  337 $\pm$ 97  &  1.48 &     0.96  &  0.59 &  1.1 &  1.85 $\pm$  0.48 &  0.207 &  0.47 & 0.31 & 0 \\
 a1407 &   10 & 0.1349&  561 $\pm$ 142 &  2.78 &      --   &  0.51 &  1.4 &  2.64 $\pm$  0.50 &  0.423 &  0.28 & 0.78 & 0 \\
 a1411 &   10 & 0.1327&  377 $\pm$ 98  &  1.41 &      --   &  0.92 &  1.1 &  1.54 $\pm$  0.31 &  0.056 &  0.67 & 0.44 & 2 \\
 a1419 &   19 & 0.1077&  504 $\pm$ 89  &  2.89 &     3.06  &  0.31 &  1.4 &  2.14 $\pm$  0.50 &  0.233 &  0.30 & 0.96 & 0 \\
 a1424 &   83 & 0.0754&  662 $\pm$ 45  &  5.49 &     4.88  &  0.14 &  1.7 &  2.32 $\pm$  0.50 &  0.476 &  0.13 & 0.13 & 0 \\
 a1437 &   33 & 0.1341&  1497 $\pm$ 13 & 39.89 &     30.5  &  0.17 &  3.2 &  7.27 $\pm$  1.24 &  3.461 &  0.08 & 0.12 & 0 \\
 a1456 &   35 & 0.1346&  540 $\pm$ 54  &  4.18 &     4.30  &  0.23 &  1.6 &  7.82 $\pm$  1.72 &  0.431 &  0.27 & 0.63 & 1 \\
 a1457 &   17 & 0.0626&  177 $\pm$ 42  &  0.29 &     0.27  &  0.61 &  0.6 &  0.40 $\pm$  0.13 &  0.000 &  0.00 & 0.29 & 2 \\
 a1468 &   49 & 0.0869&  361 $\pm$ 92  &  1.80 &     1.49  &  0.19 &  1.5 &  1.86 $\pm$  0.94 &  0.004 &  1.00 & 0.33 & 2 \\
 a1496 &   56 & 0.0958&  347 $\pm$ 46  &  1.53 &     1.29  &  0.30 &  1.1 &  1.85 $\pm$  0.38 &  0.033 &  0.50 & 0.00 & 2 \\
 a1501 &   15 & 0.1336&  406 $\pm$ 57  &  1.18 &     1.14  &  0.40 &  1.0 &  1.19 $\pm$  0.32 &  0.268 &  0.26 & 0.61 & 0 \\
 a1507 &   65 & 0.0600&  374 $\pm$ 42  &  1.36 &     0.92  &  0.23 &  1.1 &  1.47 $\pm$  0.30 &  0.072 &  0.24 & 0.76 & 0 \\
 a1516 &   72 & 0.0765&  705 $\pm$ 71  &  8.30 &     8.08  &  0.19 &  1.9 & 11.88 $\pm$  3.12 &  0.151 &  0.27 & 0.65 & 0 \\
 a1518 &   23 & 0.1065&  628 $\pm$ 118 &  4.49 &     2.40  &  0.41 &  1.6 &  2.61 $\pm$  0.85 &  0.182 &  0.23 & 0.14 & 1 \\
 a1539 &   17 & 0.1072&  510 $\pm$ 60  &  3.35 &     2.47  &  0.32 &  1.4 &  2.35 $\pm$  0.61 &  0.043 &  0.55 & 0.30 & 2 \\
 a1559 &   45 & 0.1058&  863 $\pm$ 124 & 14.06 &     11.5  &  0.33 &  2.3 &  3.61 $\pm$  0.59 &  0.193 &  0.21 & 0.02 & 0 \\
 a1564 &   57 & 0.0790&  633 $\pm$ 57  &  5.17 &     5.51  &  0.21 &  1.7 &  1.62 $\pm$  0.40 &  0.072 &  0.29 & 0.11 & 1 \\
 a1566 &   28 & 0.1015&  561 $\pm$ 69  &  3.52 &     4.04  &  0.24 &  1.5 &  1.65 $\pm$  0.40 &  0.019 &  0.67 & 0.24 & 1 \\
 a1577 &   16 & 0.1388&  359 $\pm$ 123 &  1.07 &     1.99  &  0.77 &  1.0 &  1.99 $\pm$  0.34 &  0.095 &  1.14 & 0.72 & 2 \\
 a1579 &   15 & 0.1033&  286 $\pm$ 86  &  1.00 &     1.15  &  0.73 &  1.0 &  0.46 $\pm$  0.18 &  0.033 &  0.50 & 0.23 & 1 \\
 a1581 &   16 & 0.1503&  521 $\pm$ 92  &  4.85 &     4.17  &  0.38 &  1.6 &  3.97 $\pm$  0.61 &  0.103 &  0.46 & 0.22 & 2 \\
 a1599 &   30 & 0.0855&  322 $\pm$ 38  &  0.84 &     0.58  &  0.40 &  0.9 & 10.98 $\pm$  7.96 &  3.660 &  0.09 & 0.15 & 0 \\
 a1620 &   67 & 0.0846&  782 $\pm$ 53  &  9.90 &     8.39  &  0.15 &  2.1 &  4.76 $\pm$  0.94 &  0.002 &  4.00 & 0.04 & 0 \\
 a1621 &   32 & 0.1037&  551 $\pm$ 61  &  2.12 &     2.24  &  0.31 &  1.2 & 10.03 $\pm$  7.43 &  0.088 &  0.46 & 0.03 & 0 \\
 a1646 &   27 & 0.1055&  573 $\pm$ 88  &  2.14 &     1.77  &  0.39 &  1.2 &  2.57 $\pm$  0.86 &  0.219 &  0.23 & 0.71 & 0 \\
 a1650 &   70 & 0.0839&  799 $\pm$ 87  & 11.14 &     9.51  &  0.22 &  2.1 &  4.00 $\pm$  0.75 &  3.134 &  0.06 & 0.36 & 0 \\
 a1659 &   15 & 0.1067&  383 $\pm$ 79  &  1.61 &     1.76  &  0.38 &  1.1 &  1.02 $\pm$  0.21 &  0.028 &  0.62 & 0.80 & 2 \\
 a1663 &   86 & 0.0830&  703 $\pm$ 60  &  7.62 &     7.57  &  0.17 &  1.9 &  3.01 $\pm$  0.52 &  0.548 &  0.15 & 0.22 & 0 \\
 a1674 &   17 & 0.1051&  549 $\pm$ 98  &  4.05 &     4.15  &  0.46 &  1.5 &  2.24 $\pm$  0.41 &  0.172 &  0.26 & 0.46 & 0 \\
 a1678 &   16 & 0.1689&  390 $\pm$ 124 &  1.98 &     1.64  &  0.67 &  1.2 &  1.90 $\pm$  0.50 &  0.143 &  0.50 & 0.14 & 1 \\
 a1692 &   54 & 0.0845&  561 $\pm$ 65  &  4.69 &     3.75  &  0.24 &  1.6 &  2.00 $\pm$  0.42 &  0.090 &  0.35 & 0.56 & 0 \\
 a1701 &   21 & 0.1239&  413 $\pm$ 54  &  1.15 &      1.0  &  0.49 &  1.0 &  1.76 $\pm$  0.45 &  0.138 &  0.86 & 0.49 & 1 \\
 a1750 &  115 & 0.0858&  784 $\pm$ 41  & 10.27 &     9.43  &  0.12 &  2.1 & 12.05 $\pm$  2.00 &  1.770 &  0.10 & 0.00 & 0 \\
 a1767 &  127 & 0.0705&  884 $\pm$ 55  & 11.57 &     8.68  &  0.14 &  2.2 &  3.59 $\pm$  0.64 &  1.329 &  0.05 & 0.17 & 0 \\
 a1773 &   82 & 0.0773&  779 $\pm$ 74  &  9.07 &     6.43  &  0.17 &  2.0 &  4.57 $\pm$  0.66 &  0.753 &  0.13 & 0.71 & 0 \\
 a1780 &   55 & 0.0776&  450 $\pm$ 46  &  2.51 &     2.72  &  0.22 &  1.3 &  2.55 $\pm$  0.36 &  0.033 &  0.61 & 0.17 & 1 \\
 a1809 &   99 & 0.0795&  716 $\pm$ 52  &  5.83 &     5.05  &  0.16 &  1.7 &  5.30 $\pm$  1.13 &  1.002 &  0.09 & 0.85 & 0 \\
 a1872 &   12 & 0.1480&  694 $\pm$ 138 &  3.89 &      --   &  0.43 &  1.5 &  1.95 $\pm$  0.47 &  0.253 &  0.30 & 0.05 & 0 \\
 a1882 &   55 & 0.1396&  733 $\pm$ 99  &  7.44 &     6.25  &  0.26 &  1.9 & 13.29 $\pm$  1.42 &  0.192 &  0.39 & 0.19 & 0 \\
 a1918 &   20 & 0.1402&  935 $\pm$ 129 & 16.26 &     12.3  &  0.30 &  2.4 &  5.73 $\pm$  1.68 &  2.448 &  0.08 & 0.60 & 0 \\
 a1937 &   13 & 0.1380&  223 $\pm$ 50  &  0.23 &     0.21  &  0.62 &  0.6 &  1.02 $\pm$  0.29 &  0.239 &  0.47 & 0.32 & 0 \\
 a1938 &   18 & 0.1376&  601 $\pm$ 70  &  4.95 &     5.78  &  0.26 &  1.6 &  8.82 $\pm$  3.74 &  0.714 &  0.22 & 0.52 & 0 \\
 a2026 &   51 & 0.0908&  753 $\pm$ 59  &  6.73 &     5.43  &  0.19 &  1.8 &  4.44 $\pm$  1.20 &  0.141 &  0.31 & 0.80 & 0 \\
 a2030 &   51 & 0.0915&  460 $\pm$ 54  &  2.27 &     1.80  &  0.25 &  1.3 &  2.50 $\pm$  0.40 &  0.081 &  0.35 & 0.79 & 0 \\
 a2050 &   34 & 0.1193&  826 $\pm$ 165 & 10.84 &     7.89  &  0.36 &  2.1 &  4.85 $\pm$  0.87 &  1.505 &  0.14 & 0.26 & 0 \\
 a2082 &   31 & 0.0862&  380 $\pm$ 111 &  1.84 &     1.42  &  0.56 &  1.2 &  1.99 $\pm$  0.33 &  0.065 &  0.38 & 0.56 & 0 \\
 a2094 &   36 & 0.1446&  606 $\pm$ 110 &  4.41 &     3.01  &  0.35 &  1.6 &  5.18 $\pm$  0.65 &  0.815 &  0.24 & 0.90 & 0 \\
 a2118 &   24 & 0.1416&  572 $\pm$ 91  &  4.41 &     3.34  &  0.32 &  1.6 &  3.39 $\pm$  0.91 &  0.158 &  0.37 & 0.32 & 1 \\
 a2149 &   60 & 0.0650&  330 $\pm$ 46  &  1.49 &     1.09  &  0.28 &  1.1 &  2.13 $\pm$  0.33 &  0.400 &  0.08 & 0.87 & 0 \\
 a2196 &   19 & 0.1340&  422 $\pm$ 131 &  2.17 &     2.71  &  0.58 &  1.2 &  2.72 $\pm$  0.45 &  0.492 &  0.17 & 0.80 & 0 \\
 a2211 &   15 & 0.1361&  493 $\pm$ 100 &  2.71 &     3.32  &  0.38 &  1.4 &  0.81 $\pm$  0.32 &  0.050 &  0.50 & 0.92 & 1 \\
 a2235 &   15 & 0.1492&  855 $\pm$ 195 & 12.24 &     12.2  &  0.47 &  2.2 &  6.26 $\pm$  0.96 &  1.176 &  0.14 & 0.91 & 0 \\
 a2243 &   36 & 0.1067&  759 $\pm$ 85  &  6.37 &     6.62  &  0.29 &  1.8 &  2.13 $\pm$  0.38 &  0.357 &  0.15 & 0.18 & 0 \\
 a2244 &   83 & 0.0993&  1062 $\pm$ 61 & 14.89 &     11.7  &  0.13 &  2.3 &  6.84 $\pm$  0.71 &  4.005 &  0.04 & 0.05 & 0 \\
 a2255 &  176 & 0.0801&  1121 $\pm$ 67 & 19.56 &     16.7  &  0.12 &  2.6 & 11.14 $\pm$  2.83 &  2.443 &  0.02 & 0.54 & 0 \\
 a2259 &   16 & 0.1600&  1080 $\pm$ 15 & 18.15 &     19.2  &  0.30 &  2.5 & 11.32 $\pm$  1.78 &  2.913 &  0.09 & 0.01 & 0 \\
 a2356 &   23 & 0.1195&  716 $\pm$ 85  &  5.84 &     5.45  &  0.25 &  1.7 &  3.30 $\pm$  0.59 &  0.670 &  0.18 & 0.31 & 0 \\
 a2379 &   14 & 0.1234&  531 $\pm$ 105 &  3.23 &      --   &  0.39 &  1.4 &  3.29 $\pm$  1.24 &  0.027 &  1.40 & 0.44 & 2 \\
 a2399 &  111 & 0.0579&  569 $\pm$ 37  &  4.09 &     3.19  &  0.14 &  1.5 &  3.54 $\pm$  0.43 &  0.490 &  0.12 & 0.07 & 0 \\
 a2428 &   42 & 0.0839&  420 $\pm$ 23  &  2.17 &     2.12  &  0.15 &  1.2 &  2.57 $\pm$  0.67 &  1.351 &  0.14 & 0.84 & 0 \\
 a2433 &   16 & 0.1195&  257 $\pm$ 44  &  0.50 &     0.33  &  0.40 &  0.8 &  0.84 $\pm$  0.22 &  0.092 &  0.47 & 0.66 & 0 \\
 a2448 &   38 & 0.0820&  447 $\pm$ 62  &  2.56 &     2.14  &  0.28 &  1.3 &  2.05 $\pm$  0.49 &  0.029 &  0.54 & 0.64 & 2 \\
 a2505 &   21 & 0.1100&  366 $\pm$ 57  &  0.94 &     1.01  &  0.40 &  1.0 &  1.58 $\pm$  0.29 &  0.238 &  0.28 & 0.00 & 0 \\
 a2561 &   13 & 0.1634&  405 $\pm$ 91  &  1.17 &     1.23  &  0.50 &  1.0 &  2.16 $\pm$  0.38 &  1.720 &  0.25 & 0.72 & 1 \\
 a2564 &   20 & 0.0828&  339 $\pm$ 70  &  1.25 &     1.40  &  0.45 &  1.0 &  0.93 $\pm$  0.20 &  0.041 &  0.38 & 0.45 & 0 \\
 a2593 &  167 & 0.0419&  570 $\pm$ 55  &  4.53 &     3.68  &  0.18 &  1.6 &  3.24 $\pm$  0.53 &  0.485 &  0.07 & 0.74 & 0 \\
 a2670 &  109 & 0.0761&  804 $\pm$ 51  &  9.40 &     9.32  &  0.13 &  2.0 &  5.08 $\pm$  0.56 &  1.255 &  0.10 & 0.73 & 0 \\
 a2705 &   33 & 0.1165&  452 $\pm$ 64  &  3.45 &     3.65  &  0.30 &  1.5 &  3.77 $\pm$  0.58 &  0.018 &  0.50 & 0.03 & 1 \\
\hline
\hline
\end{longtable}

\end{appendix}


\begin{thebibliography}{}
\bibitem[Abell (1958)]{abell1}
Abell, G.O. 1958, ApJ, 3, 211
\bibitem[Abell (1958)]{abell2}
Abell, G. O., Corwin, H. G. Jr., Olowin, R. P. 1989, ApJ, 70, 1
\bibitem[Abazajian et al.(2003)]{dr1}
Abazajian, K., Adelman, J., Agueros, M.,et al. 2003, AJ, 126, 2081 (Data Release One)
\bibitem[Adami et al. (1998)]{abm}
Adami, C., Biviano, A., \& Mazure, A. 1998c, A\&A, 331, 439
\bibitem[Adami et al.(1998a)]{adami} 
Adami, C., Mazure, A., Biviano, A., Katgert, P., \& Rhee, G. 1998a, A\&A, 331, 493
\bibitem[Adami et al.(1998b)]{adamib} 
Adami, C., Mazure, A., Katgert, P., \& Biviano, A. 1998b, A\&A, 336, 63
\bibitem[Akritas et al.(1996)]{akritas}
Akritas, M. G., Bershady, M. A. 1996, ApJ, 470, 706
\bibitem[Bahcall et al. (2003)]{bahcom2} 
Bahcall, N.A., McKay, T.A,, Annis, J., et al.  2003, ApJ, 148, 243
\bibitem[Barrena et al. (2002)]{barre}
Barrena, R., Biviano, A., Ramella, M., Falco, E.E., \& Seitz, S. 2002, 
A\&A, 386, 816
\bibitem[Bartelmann (1996)]
Bartelmann, M. 1996, A\&A, 313, 697
\bibitem[Basilakos et al. (2004)]{basi}
Basilakos, S., Plionis, M., Georgakakis, A., Georgantopoulos, I.  2004, MNRAS, 351, 989
\bibitem[Beers et al.(1990)]{beers2}
Beers, T.C., Flynn,K., Gebhardt 1990, AJ, 100, 32
\bibitem[Beers et al.(1991)]{beers3}
Beers, T.C., Forman, W., Huchra, J.P., Jones, C., \& Gebhardt, K.
1991, AJ, 102, 1581
\bibitem[Biviano \& Katgert (2004)]{bivkat}
Biviano, A., \& Katgert, P. 2004, A\&A, 424, 779
\bibitem[Biviano et al.(1997)]{biviano1}
Biviano, A., Katgert, P., Mazure, A. et al. 1997, A\&A, 321, 84
\bibitem[Biviano et al.(2002)]{biviano2}
Biviano, A., Katgert, P., Thomas, T. et al.  2002, A\&A, 387, 8
\bibitem[Blanton et al.(2003)]{blanton}
Blanton, M.R., Lupton, R.H., Maley, F.M. et al. 2003, AJ, 125, 2276
\bibitem[Blanton et al.(2005)]{blanton}
Blanton, M.R., Lupton, R.H., Schlegel, D.J. et al., ApJ, 631, 208
\bibitem[B\"ohringer et al.(2000)]{bh1}
B\"ohringer, H., Voges, W.; Huchra, J. P., et al. 2000, ApJS, 129, 435
\bibitem[B\"ohringer et al.(2001)]{bh2}
B\"ohringer, H.,  Schuecker, P., Guzzo, L., et al. 2001, A\&A, 369, 826
\bibitem[B\"ohringer et al.(2002)]{bh3}
B\"ohringer, H., Collins, C. A., Guzzo, L., et al. 2002, ApJ, 566, 93
\bibitem[Borgani \& Guzzo (2001)]{borgans0}
Borgani S., \& Guzzo, L.  2001, Nature, 409, 39
\bibitem[Bower et al.(1994)]{bower}
Bower, R. G., Bohringer, H., Briel, U. G., et al.  1994, MNRAS, 268, 345
\bibitem[Bower et al.(1997)]{bower1}
Bower, R. G., Castander, F. J., Ellis, R. S., et al.  1997, MNRAS, 291, 353
\bibitem[Carlberg et al.(1997)]{carl1}
Carlberg, R. G., Yee, H. K., Ellingson, E. 1997a, ApJ, 478, 462 
\bibitem[Castander et al.(1994)]{casta}
Castander, F.J., Ellis, R.S., Frenk, C.S., Dressler, A., Gunn, J.E. ApJ, 1994, 424, 79
\bibitem[Castander et al.(1995)]{casta1}
Castander, F. J., Bower, R. G., Ellis, R. S., 1995, Nature, 377, 39
\bibitem[Castander et al.(1995)]{casta}
Castander, F. J., Bower, R. G., Ellis, R. S., 1995, Nature, 377, 39
\bibitem[Clowe et al. (2004)]{clowe}
Clowe, D. , De Lucia, G., \& King, L. 2004, MNRAS, 350, 1038
\bibitem[Couch et al.(1991)]{couch}
Couch, W. J., Ellis, R. S., MacLaren, I., Malin, D. F. 1991, MNRAS, 249, 606
\bibitem[Dalton et al.(1994)]{Dalton}
Dalton G.B., Efstathiou G., Maddox S.J., Sutherland W.J. 1994, MNRAS, 269, 151
\bibitem[den Hartog \& Katgert 1996]{denhart}
den Hartog, R., \& Katgert, P. 1996, MNRAS, 279, 349
\bibitem[D\'{\i}az et al. 2005]{diax}
D\'{\i}az, E., Zandivarez, A., Merch\'an, M.E., \& Muriel, H. 2005, ApJ, 629, 158
\bibitem[Dolag et al. (2004)]{dola} 
Dolag, K., Bartelmann, M., Perrotta, F., Baccigalupi, C., Moscardini, L., Meneghetti, M., \& Tormen, G. 2004, A\&A, 416, 853
\bibitem[Donahue \& Voigt (1999)]{dona0}
Donahue, M., \& Voigt, M. 1999, ApJ, 523, 137
\bibitem[Donahue et al. (2002)]{dona}
Donahue, M., Scharf, C.A., Mack, J. et al. 2002, ApJ, 569, 689
\bibitem[Dressler (1980)]{dress80}
Dressler, A. 1980, ApJ, 236, 351
\bibitem[Dressler \& Shectman (1988)]{dress}
Dressler, A. \& Shectman, S. A. 1988, AJ, 95, 985
\bibitem[Ebeling et al. (1996a)]{eb1}
Ebeling, H., Voges, W., Bohringer, H., et al. 1996a, MNRAS, 283, 1103
\bibitem[Ebeling et al. (1996b)]{eb2}
Ebeling, H., Voges, W., Bohringer, H., et al. 1996b, MNRAS, 281, 799
\bibitem[Ebeling et al. (2000)]{eb4}
Ebeling, H., Edge A.C., Allen S.W., Crawford C. S., Fabian A.C., Huchra, J.P. 2000, MNRAS, 318, 333
\bibitem[Eisenstein et al.(2001)]{eis}
Eisenstein, D.J., Annis, J., Gunn, J.E., et al. 2001, AJ, 122, 2267
\bibitem[Eke et al.(1996)]{eke}
Eke, V. R., Cole, S., Frenk, C. S. et al. 1996, MNRAS, 282, 263
\bibitem[Fukugita et al.(1996)]{fuk2}
Fukugita, M., Ichikawa, T., Gunn, J. E. 1996, AJ, 111, 1748
\bibitem[Gilbank et al. (2004)]{gilbank}
Gilbank, D.G., Bower, R.G., Castander, F.J., \& Ziegler, B.L. 2004, MNRAS, 348, 551 
\bibitem[Gioia et al. (2001)]{gioia}
Gioia, I. M., Henry, J. P., Mullis, C. R. et al. 2001, ApJ, 553, 105
\bibitem[Girardi et al.(1993)]{marisa1}
Girardi, M., Biviano, A., Giuricin, G. et al. 1993, ApJ, 404, 38
\bibitem[Girardi et al.(1998)]{marisa3}
Girardi, M., Giuricin, G., Mardirossian F., Mezzetti, M. et al. 1998, ApJ, 505, 74
\bibitem[Gladders et al.(2000)]{glad}
Gladders, M. D., Yee, H. K. C. 2000, AJ, 120, 2148
\bibitem[Goto et al.(2002)]{goto}
 Goto, T., Sekiguchi, M., Nichol, R. C. et al. 2002, AJ, 123, 1807
\bibitem[Gunn et al.(1996)]{gunn0}
Gunn, J. E., Hoessel, J. G., Oke, J. B. 1986, ApJ, 306, 30
\bibitem[Gunn et al.(1998)]{gunn}
Gunn, J.E., Carr, M.A., Rockosi, C.M., et al 1998, AJ, 116, 3040 (SDSS Camera)
\bibitem[Holden et al.(1997)]{Holden}
Holden B.P., Romer A.K., Nichol R.C., Ulmer M.P. 1997, AJ, 114, 1701
\bibitem[Hogg et al.(2001)]{Hogg}
Hogg, D.W., Finkbeiner, D. P., Schlegel, D. J., Gunn, J. E. 2001, AJ, 122, 2129
\bibitem[Horner(2001)]{horner}
Horner, D. 2001, PhD Thesis, University of Maryland
\bibitem[Katgert et al.(1996)]{kat96}
Katgert, P., Mazure, A., Perea, J., er al. 1996, A\&A 310, 8
\bibitem[Katgert et al.(2004)]{kat04}
Katgert, P., Biviano, A., \& Mazure, A.  2004, ApJ, 600, 657
\bibitem[King (1966)]{kin66} 
King, I. R. 1966, AJ, 71, 64
\bibitem[Ledlow et al.(2003)]{ledlow}
Ledlow M.J., Voges W., Owen F.N., Burns J.O. 2003, AJ, 126, 2740
\bibitem[Lubin et al.(2004)]{lubin}
Lubin L.M., Mulchaey J.S., Postman M. 2004, ApJ, 601, 9
\bibitem[Lumsden et al.(1992)]{lumsden}
Lumsden, S. L.,  Collins, C. A., Nichol, R. C.,et al. 1992, MNRAS, 258, 1
\bibitem[Lupton et al.(1999)]{lup1}
Lupton, R. H., Gunn, J. E., Szalay, A. S. 1999, AJ, 118, 1406
\bibitem[Lupton et al.(2001)]{lup2}
Lupton, R., Gunn, J. E., Ivezi\'c, Z.,  et al.  2001, in ASP Conf. Ser. 238, Astronomical Data Analysis Software and Systems X, ed. F. R. Harnden, Jr., F. A. Primini, and H. E. Payne (San Francisco: Astr. Soc. Pac.), p. 269 (astro-ph/0101420)
\bibitem[Mazure (1996)]{mazure}
Mazure, A., Katgert, P., den Hartog, R. et al. 1996, A\&A, 310, 31
\bibitem[Merritt (1987)]{merritt}
Merritt, D.R. 1987, ApJ, 313, 121
\bibitem[Mulchaey et al.(2003)]{mul}
Mulchaey, J.S., Davis, D. S., Mushotzky, R. F.; Burstein, D. 2003,ApJS, 145, 39
\bibitem[Navarro et al.(1996)]{nav1}
Navarro, J. F., Frenk, C. S., White, S. D.M. 1996, ApJ, 462, 563
\bibitem[Navarro et al.(1997)]{nav2}
Navarro, J. F., Frenk, C. S., White, S. D.M. 1997, ApJ, 490, 493
\bibitem[Olsen et al.(1999)]{olsen}
Olsen, L. F., Scodeggio, M., da Costa, L. et al. 1999 A\&A, 345, 681
\bibitem[Popesso et al.(2005)]{pop}
Popesso, P., B\"ohringer, H., Brinkmann J., et al. 2004, A\&A, 423, 449, Paper I
\bibitem[Popesso et al.(2005a)]{pop1}
Popesso, P., B\"ohringer, H., Romaniello, M., \& Voges, W. 2005b, A\&A, 433, 415, Paper II
\bibitem[Popesso et al.(2005b)]{pop2}
Popesso, P., A. Biviano, B\"ohringer, H., Romaniello, M. 2005a, A\&A, 433, 431, Paper III
\bibitem[Popesso et al.(2005c)]{pop3}
Popesso, P., A. Biviano, B\"ohringer, H., Romaniello, M., \& Voges, W. 2005c, astro-ph/0506201, Paper IV
\bibitem[Postman et al.(1996)]{post}
Postman, M., Lubin, L. M., Gunn, J. E., et al. 1996, AJ, 111, 615
Retzlaff, J. 2001,XXIst Moriond Astrophysics Meeting, March 10-17, 2001 Savoie, France. Edited by D.M. Neumann  J.T.T. Van.
\bibitem[Rosati, Borgani \& Norman (2002)]{rosa}
Rosati, P., Borgani, S., Norman, C. 2002, ARA\&A, 40, 539
\bibitem[Scharf et al.(1997)]{scharf}
Scharf, C. A., Jones, L. R., Ebeling, H., et al.  1997, ApJ, 477, 79
\bibitem[Smith at al.(2002)]{smith}
Smith, J.A., Tucker, D.L., Kent, S.M., et al. 2002, AJ, 123, 2121
\bibitem[Stocke at al.(1991)]{sto}
Stocke, J.T., Morris, S. L., Gioia, I. M., et al.  1991, ApJ, 76, 813
\bibitem[Stoughton et al.(2002)]{stoughton}
Stoughton, C., Lupton, R.H., Bernardi, M., et al. 2002, AJ, 123, 485
\bibitem[Strateva et al.(2001)]{strateva}
Strateva, I., Ivezi\'c, Z., Knapp, G., et al. 2001 AJ, 122, 1861
\bibitem[Strauss et al.(2002)]{strauss}
Strauss, M. A., M.A., Weinberg, D.H., Lupton, R.H. et al. 2002, AJ, 124, 1810
\bibitem[The \& White (1986)]{the}
The, L. S., White, S. D. M.  1986, AJ, 92, 1248
\bibitem[trumper (1988)]{tru}
Truemper, J. 1988, NATO Advanced Science Institutes (ASI) Series C, 249, 355
\bibitem[van der Marel et al. (2000)]{vdm00}
van der Marel, R.P., Magorrian, J., Carlberg, R.G., Yee, H.K.C., \&
Ellingson, E. 2000, AJ, 119, 2038
\bibitem[Voges et al. (1999)]{voges}
Voges, W., Aschenbach, B., Boller, T. et al. 1999, A\&A, 349, 389
\bibitem[Wojtak et al. (2005)]{woj}
Wojtak, R., \L okas, E.L., Gottl\"ober, S., \& Mamon, G.A. 2005, MNRAS, 361, L1
\bibitem[Yasuda et al.(2001)]{yasuda}
Yasuda, N., Fukugita, M. Narayanan, V. K. et al. 2001, AJ, 122, 1104
\bibitem[York et al.(2000)]{york}
York, D. G., Adelman, J., Anderson, J.E.,  et al. 2000, AJ, 120, 1579
\bibitem[Zwicky et al (1968)]{zwi2}
Zwicky, F., Herzog, E., Wild, P., Karpowicz, M., \& Kowal, C.
1961--1968, Catalog of Galaxies and Clusters of Galaxies 1--6
\end{thebibliography}
\end{document}